%% file: paper.tex
\documentclass[acmtog, authorversion, review=false, nonacm, timestamp]{acmart}

\setcopyright{acmlicensed}\acmJournal{TOG}

\input{preamble.tex}

\def\teasersize{0.975}
\def\teasercaptionoffset{-2mm}
\setcopyright{none}
\renewcommand\footnotetextcopyrightpermission[1]{} 
\def\imagesmode{}
\def\imagesdir/{images/}
\def\preprint{1}

\if\preprint1
    \renewcommand\footnotetextcopyrightpermission[1]{} 
    \setcopyright{none}
    \AtBeginDocument{%
        \addtolength{\footskip}{2.0\baselineskip}%
        \fancyfoot[LE]{\textit{\textbf{Preprint}}}%
        \fancyfoot[RO]{\textit{\textbf{Preprint}}}%
    }
\fi

\newtcolorbox{gtcb}[4]{
    enhanced,
    boxed title style={top=0pt, bottom=0pt, arc=.75mm, boxrule=0.2mm, colframe=white},
    attach boxed title to top left={xshift=-3mm, yshift*=-\tcboxedtitleheight/2},
    left=0pt,
    right=0pt,
    top=0pt,
    bottom=.5mm,
    boxrule=0mm,toprule=.125mm,bottomrule=0mm,leftrule=.125mm,left=1mm,right=1mm,
    titlerule=0mm,toptitle=0mm,bottomtitle=0mm,top=-1mm,
    colframe=gray,
    colback=#1,
    coltitle=#2,
    colbacktitle=#3,
    title=#4,
    fonttitle=\bfseries\sffamily\footnotesize,fontupper=\normalsize\itshape,
}

\DeclareMathOperator\timereversal{-\mkern-.5mu t}

\input{fig_teaser.tex}

\begin{document}


\externaldocument{supplemental}


\title[A Generalized Ray Formulation For Wave-Optics Rendering]%
      {A Generalized Ray Formulation For Wave-Optics Rendering}

\author{Shlomi Steinberg}
\email{p@shlomisteinberg.com}
\orcid{0000-0003-2748-4036}
\additionalaffiliation{%
    \institution{University of Waterloo}
    \state{Ontario} 
    \country{Canada}
}

\author{Ravi Ramamoorthi}
\email{ravir@cs.ucsd.edu}
\orcid{0000-0003-3993-5789}
\additionalaffiliation{%
    \institution{University of California San Diego}
    \state{California}
    \country{United States}
}

\author{Benedikt Bitterli}
\email{benedikt.bitterli@gmail.com}
\orcid{0000-0002-8799-7119}

\affiliation{%
    \institution{NVIDIA}
    \city{San Francisco}
    \country{United States}
}

\author{Eugene d'Eon}
\email{ejdeon@gmail.com}
\orcid{0000-0002-3761-2989}

\affiliation{%
    \institution{NVIDIA}
    \city{Wellington}
    \country{New Zealand}
}

\author{Ling-Qi Yan}
\email{lingqi@cs.ucsb.edu}

\affiliation{%
    \institution{University of California, Santa Barbara}
    \city{Santa Barbara}
    \state{California}
    \postcode{93106}
    \country{United States}
}

\author{Matt Pharr}
\email{matt@pharr.org}
\orcid{0000-0002-0566-8291}

\affiliation{%
    \institution{NVIDIA}
    \city{San Francisco}
    \country{United States}
}

\begin{abstract}
    Under ray-optical light transport, the classical ray serves as a linear and local ``point query'' of light's behaviour.
    Linearity and locality are crucial to the formulation of sophisticated path tracing and sampling techniques, that enable efficient solutions to light transport problems in complex, real-world settings and environments.
    However, such formulations are firmly confined to the realm of ray optics, while many applications of interest---in computer graphics and computational optics---demand a more precise understanding of light: as waves.
    We rigorously formulate the \emph{generalized ray}, which enables linear and weakly-local queries of arbitrary wave-optical distributions of light.
    Generalized rays arise from photodetection states, and therefore allow performing backward (sensor-to-source) wave-optical light transport.
    Our formulations are accurate and highly general:
    they facilitate the application of modern path tracing techniques for wave-optical rendering, with light of any state of coherence and any spectral properties. 
    We improve upon the state-of-the-art in terms of the generality and accuracy of the formalism, ease of application, as well as performance.
    As a consequence, we are able to render large, complex scenes, as in \cref{fig_teaser}, and even do interactive wave-optical light transport, none of which is possible with any existing method.
    We numerically validate our formalism, and make connection to partially-coherent light transport.
\end{abstract}




\maketitle

\if\preprint1
    \thispagestyle{plain}
\fi



\section{Introduction} \label{section_introduction_paper}

Rendering and light transport theories are often formulated within the context of \emph{ray optics}.
We may state three tenets of ray-optical light transport:
\begin{enumerate}
    \item   \textbf{Locality}: 
            The light-transport primitive, i.e. the \emph{ray}, exists at a singular position (a point) in space and propagates into a singular direction at any given time.
            Therefore, rays serve as perfect ``point queries'' of light's behaviour. 
    \item   \textbf{Linearity} in terms of radiometric quantities (like intensity or radiance): 
            The observed intensity $I$ of a superposition of a pair of rays, with intensities $I_1,I_2$, is 
            \begin{align}
                I = I_1 + I_2
                \label{ray_superposition}
                ~.
            \end{align}
    \item   \textbf{Completeness}: 
            Any observable optical phenomenon (within the ray-optical context) can be computed to arbitrarily high accuracy with a finite count of rays.
\end{enumerate}
The first pair of tenets are essential for rendering and path tracing techniques.
\emph{Locality} enables the use of spatial-subdivision acceleration structures, facilitating efficient ray propagation, even in the presence of massive geometry and detail.
\emph{Linearity} gives rise to a linear rendering equation,
enabling the application of powerful sampling techniques. 
Note that linearity must be formulated in terms of radiometric quantities, as opposed to, for example, field strength.
The final tenet, \emph{completeness}, makes these light transport theories general: any effect that can be described by the applicable physics (ray optics) may also be simulated.

The locality and linearity of ray-optical light transport allows these techniques to scale to the massive scenes that can be seen in computer-generated content in films and other media.
Today, path tracing techniques are even applied in real-time applications using dedicated ray-tracing hardware \cite{Burgess2020-eq}.
The above comes at a price: ray optics is a simplistic understanding of light, where wave effects are ignored.
Such effects include the colourful glints that appear when light is scattered by scratches in metal; the colour of the wings and scales of some species of insects, snakes and fish; stress birefringence; and, the appearance of an oil layer or metal oxide on a surface.
Many important unsolved problems demand wave simulation in complex scenes, e.g., RADAR simulation in urban environments for automotive applications, and signal coverage simulation for efficient infrastructure deployment.

The appearance of diffractive materials (\cref{fig_teaser}), the ability of long-wavelength RADAR radiation to diffract around an object, and other wave effects all depend on light's wave properties, its spatial and temporal optical coherence and spectral properties.
In other words, the applications and effects mentioned above cannot be simulated only at a material-level, e.g., via a ``diffractive BRDF'' as done in computer graphics.
Instead, these phenomena are the aggregated product of the \emph{wave-optical light transport} over the entire scene. 
In \cref{fig_snake} we show how light's wave properties induce very different appearance in the same materials.
Our focus in this paper is on formulating a wave-optical light transport formalism, that allows efficient and accurate modelling of these effects.

One could be na\"ively motivated to attempt to formulate a wave-optical formalism that would admit all the three important tenets above.
That is, replace the wave function with an alternative analytic construct, that is perfectly local, linear in terms of its observed intensity, and wave-optically complete.
Unfortunately, that is impossible: a single-point formulation of light---one that is simultaneously both \emph{local and linear}---is incompatible with electromagnetism \cite{wolf2007introduction}.

While locality and linearity cannot be simultaneously achieved, many computational electromagnetic frameworks have been designed to be perfectly local: i.e., they employ a ray-like construct to conduct their wave simulations.
This is indeed possible, and such formulations may even be complete, however any such formalism inevitably abandons linearity.
This means that the observed intensity of a superposition of these ray-like constructs does not follow \cref{ray_superposition}, but instead is:
\begin{align}
    I = I_1+I_2+\mathfrak{I}_{12}
    \label{bilinear_superposition}
    ~,
\end{align}
where $\mathfrak{I}_{12}$ is a \emph{bilinear} interference term that depends on both rays.
The loss of linearity greatly frustrates the application of sampling and path tracing techniques: clearly, a linear rendering equation cannot be formulated; and the angular distribution of scattered energy---the BRDF---at a point of interaction behaves non-linearly, thus practical importance sampling is a significant challenge.

Linearity is crucial for a formulation that is compatible with path tracing techniques, hence we conclude that perfect locality must be abandoned.
Recognizing that fact, \emph{physical light transport} (PLT) was introduced by \citet{Steinberg_practical_plt_2022,Steinberg_lt_framework_2021} as a framework that enables practical, \emph{weakly-local} wave-optical path tracing. 
PLT forgoes locality \emph{to a degree}, that is, in-place of rays, PLT's light transport primitives are beams:
While a ray, at any time instant, exists at a singular point, a beam occupies a small spatial region and propagates into a solid angle.
The important insight here is that, under some circumstances, locality can be sacrificed to a partial degree, in order to regain perfect linearity.

PLT deals with partially-coherent light, and sets the spatial region occupied by a beam to be proportional to the spatial coherence of light.
Thereby, distinct beams become mutually incoherent, the interference term in \cref{bilinear_superposition} vanishes, and beams always superpose linearly.
Hence, PLT is able to formulate a \emph{linear} wave-optical rendering equation \cite{Steinberg_lt_framework_2021}, 
enabling the application of some classical path tracing tools and importance sampling of light-matter interactions.
PLT is able to scale to large scenes, beyond any non-linear framework, while remaining accurate within its applicability domain (weakly-coherent light).

\paragraph{The sampling problem}

In order to recover weak locality, PLT builds upon the partial coherence of light to formulate mutually-incoherent beams.
The implication is that PLT is only compatible with a \emph{forward} model of light transport, i.e. where we trace light from light sources to a sensor.
To see why, recall that the coherence properties of light depend upon the light source, mutate on propagation, and on interaction with matter.
The spatial extent of a beam, and hence \emph{the degree-of-locality of the formalism}, also depends on these properties.
If we were to trace beams \emph{backwards} (from a sensor to sources) we would not be able to localize the beam, as the coherence properties of light are unknown, therefore its spatial extent may be arbitrary, and locality is lost in its entirety (see \cref{fig_sampling_problem}).

We call this incompatibility of weakly-local formalisms with backward transport the \emph{sampling problem}.
The sampling problem is not restricted to PLT only, but applies to other formalisms as well, as we will discuss in \cref{section_related_work}. 
Moreover, in a forward model of light transport, weak locality can only be recovered by assuming or enforcing constraints on light's properties, therefore:
\begin{enuminline}
    \item
        any forward weakly-local formalism suffers from the sampling problem;
    \item 
        and, such formalisms are never complete.
\end{enuminline}

\input{fig_sampling_problem.tex}

\paragraph{Our contributions in this paper}

The backward model plays an important practical role in light transport simulations (for example, modern path tracers trace backward only or bi-directionally), and our primary motivation is to solve the sampling problem.
In sharp contrast to ray optics, where a ray of light admits identical dynamics to its time-reversed counterpart, within the wave-optical context there is a fundamental change in physics between forward and backward light transport.
We will discuss this further in \cref{section_theory}.

The sampling problem arises due to this fundamental incompatibility between the forward and backward models.
In order to properly address this sampling problem, we derive a novel formalism of backward wave-optical light transport.
Our formalism is simultaneously \emph{weakly-local, linear and complete}. 
There does not exist another wave-optical formalism with all these properties, and, as discussed, this would not be possible in a forward model.
This is a powerful result: completeness means that our formalism can sample arbitrary wave-optical distributions of light, of any wavelength, any spectrum, any polarization, and any optical coherence.

At the heart of our proposed formalism is the \emph{generalized ray}: a general-purpose wave-optical construct, that generalizes the classical ``point query'', i.e. the ray, to wave optics.
Conveniently, the weak locality of a generalized ray does not dependent on light's properties, and only depends on the sensed wavelength.
In \cref{section_backward_light_transport,section_gr_analysis}, we derive our theory of backward wave-optical light transport, making essentially no approximating assumptions, from well-known optical principles.
This is our primary contribution.
The sections marked with an asterisk (\hardsec) are mathematically dense, and readers that are less interested in the theory might wish to (initially) skip these sections.

While our proposed formalism may also serve as an alternative to the state-of-the-art (for backward-only transport), it is meant to \emph{complement} existing weakly-local, linear forward-based formulations, by working in tandem with them to enable bi-directional wave-optical transport.
In \cref{section_sample_solve} we discuss a particular application of the formalism to high-performance wave-optical rendering, and present \emph{sample-solve}: a simple two-stage bi-directional algorithm, that first samples the scene via backward transport with generalized rays; then, uses PLT to apply a partially-coherent forward pass that acts as a variance reduction technique.
We provide a sample implementation of our rendering algorithm in our supplemental material.
We will show in \cref{section_interactive_rendering} that we are able to do wave-optical rendering with complex scenes (as in \cref{fig_teaser}), and at interactive performance, with convergence rate that is multiple orders-of-magnitude faster compared to the state-of-the-art.


\section{Related Work} \label{section_related_work}

\paragraph{Wave-optical light transport}

In computer graphics one of the earliest formulations of wave-optical light transport is \emph{diffractive shaders} \cite{Stam_1999}.
Light is propagated as rays, and a ray is converted into a plane wave (implicitly an asymptotic, far-field approximation) when formulating a diffractive shader, which quantifies the interaction of that ray with a surface.
A plane wave exists throughout the entire space, 
therefore no degree of locality could be rigorously recovered.

Wigner distribution-based light transport methods have gained some attention both in computer graphics \cite{Cuypers_Haber_Bekaert_Oh_Raskar_2012} and optical literature \cite{mout2018ray,mackay2021millimetre,jensen1991methodology}.
The \emph{Wigner distribution function} (WDF) $\wvd(\va{r},\va{k})$ (see formal definition in \cref{def_WDF}) is a bilinear distribution defined w.r.t. some wave function $\psi(\va{r})$, and is a function of spatial position $\va{r}$ and wavevector $\va{k}$. 
The WDF admits some useful ray-like properties, which has motivated the definition of the WDF as a form of a ``generalized radiance'' \cite{walther1968radiometry}.
However, locality is lost when the interaction with matter is considered, as we formally show next.
The scattered WDF is \cite[Chapter 8]{torre2005linear}
\begin{align}
    \wvd[o]\qty(\va{r}_o,\va{k}_o) =& 
        \int 
            \dd{\va{r}_i} \dd{\va{k}_i} 
            K\qty(\va{r}_o,\va{r}_i,\va{k}_o,\va{k}_i) 
            \wvd[i]\qty(\va{r}_i,\va{k}_i)
    \label{wdf_convolution_kernel}
    ~,
\end{align}
where $\wvd[i],\wvd[o]$ are the input and output WDFs, respectively, and $K$ is a \emph{diffraction kernel} that is completely defined by the scattering matter \cite[Chapter 1.6]{testorf2010phase}:
\begin{align}
    \!\!\!
    K\qty(\va{r}_o,\va{r}_i,\va{k}_o,\va{k}_i) 
    \triangleq&
        \int \dd{\va{x}_o} \dd{\va{x}_i} 
            h\qty(\va{r}_o+\tfrac{1}{2}\va{x}_o,\va{r}_i+\tfrac{1}{2}\va{x}_i)
    \nonumber \\[-.1ex]
    &\times
            h^\star\qty(\va{r}_o-\tfrac{1}{2}\va{x}_o,\va{r}_i-\tfrac{1}{2}\va{x}_i)
            \ee^{-\ii \qty(\va{k}_o\cdot\va{x}_o - \va{k}_i\cdot\va{x}_i)}
    \label{WDF_kernel_def}
    ~,
\end{align}
where $\star$ denotes complex conjugation, and $h$ is the \emph{optical response function} of the system.
Locality is lost because, by definition, $K$ requires integration over the \emph{entire scene}.

\citet{Cuypers_Haber_Bekaert_Oh_Raskar_2012} claim that the negative values taken by the WDF allow ``rays to interfere later for global illumination''.
This is only true when a single WDF accounts for all light in the system, and the kernel $K$ accounts for the entire scene.
The WDF is a bilinear distribution, therefore the WDF of the sum $\psi_1+\psi_2$ takes a form identical to \cref{bilinear_superposition}, viz.
\begin{align}
    \wvd = \wvd[1] + \wvd[2] + \wvd[12]
    \label{wdf_bilinear}
    ~,
\end{align}
where $\wvd[1],\wvd[2]$ are the WDFs of $\psi_1,\psi_2$, respectively, and $\wvd[12]$ is a bilinear cross-term, that depends on both $\psi_1$ and $\psi_2$ (see \citet[Chapter 1.8]{testorf2010phase} for more details).
If we were to limit the spatial integration in \cref{WDF_kernel_def} to a finite region, then different kernels would account for different interactions, and to compute the total superposition WDF with contributions from rays scattered from distinct spatial regions, we must account for the bilinear term in \cref{wdf_bilinear}.
\citet{Cuypers_Haber_Bekaert_Oh_Raskar_2012} implicitly neglect that term: by assuming that the WDFs from different rays add up \emph{linearly}.
The negative values that arise in the WDF of a light distribution depend only on that distribution, while the bilinear cross-term that integrates over both distributions is the term that accounts for their mutual interference.
Discarding the cross-term is equivalent to neglecting wave interference.
We numerically demonstrate in \cref{fig_twin_slit_cuypers_plots} that limiting the integration region in \cref{WDF_kernel_def} and not accounting for the bilinearity of the WDF produces incorrect results.

Both \citet{Stam_1999} and \citet{Cuypers_Haber_Bekaert_Oh_Raskar_2012} recognize the need for linearity, however neither admits machinery that is able to limit the spatial integration region of light-matter interactions to a well-defined finite extent, hence neither admits locality of any form.
Even if such machinery was to be developed, these frameworks would then suffer (like any other weakly-local forward formalisms) from the very same sampling problem that we address in this paper.

Physical light transport (PLT) \cite{Steinberg_lt_framework_2021,Steinberg_practical_plt_2022} does derive such machinery by quantifying the coherence of light.
Under the assumption of weakly-coherent light, rigorous, well-defined weak locality is recovered.
That is, PLT is able to quantify the spatial extent over which a light-matter interaction (like \cref{WDF_kernel_def}) may produce observable interference.
Treating light as partially-coherent beams, PLT gives rise to a weakly-local, linear formalism of wave-optical light transport.
PLT is able to accurately, and efficiently, reproduce partially-coherent effects.
However, in order to do that, PLT needs to propagate the coherence properties of light forward---properties that are needed in-order to recover weak locality---hence the sampling problem inevitably arises, and practical backward light transport is not possible.

\paragraph{Computational electrodynamics}

A variety of wave solvers, many commercially available, aim to produce a solution to Maxwell's equations.
Common solvers are variants of the finite-difference time-domain (FDTD) \citep{yee1966numerical}, finite element (FEM) \cite{jin2015finite}, or boundary element (BEM) methods.
These methods are usually supra-linear in scene complexity, require deterministically described fields as well as geometry at sub-wavelength resolution.
Furthermore, solving for the electromagnetic field directly is neither feasible nor desirable in most applications of interest: \emph{observable properties} of light arise when integrated over the extent of a sensor as well as over the period of observation.
These methods are practical only in exceedingly simple scenes.

Of more relevance are ``physical optics'' methods, most notably the ``shooting-bouncing ray'' (SBR) method.
SBR is a perfectly-local method, by design, and uses ray tracing to track field propagation to a target aperture or geometry (i.e., a ray captures the behaviour of the field as it propagates into some solid angle).
Then, the electromagnetic field is reconstructed at the target in order to evaluate the scattered electromagnetic radiation.
Applications of SBR methods are extensive, and include:
simulation of radar for imaging \cite{feng2021multiview}; driving-assistive technology \cite{Castro_Singh_Arora_Louie_Senic_2019}; 
analysis of aircraft scattering cross-section \cite{bilal2019comparison}; ground-penetrating radar \cite{Warren_Giannopoulos_Giannakis_2016}; 
and, indoor positioning using WiFi \cite{hossain2018indoor}.
Some methods point-sample an input field and propagate a ray tube in order to approximate a diffraction integral \cite{Andreas2015}.

Another popular class of perfectly-local methods are ray tracers that handle diffractions by employing the geometric/uniform theory of diffraction (UTD) \citep{son1999deterministic,bilibashi2020dynamic,yi2022ray}.
Some of these have become the state-of-the-art in the simulation of light of longer wavelengths: major applications include automotive-targeted simulation of RADAR \citep{boban2014geometry,guan2020through}, as well as simulation of WiFi/cellular radiation (e.g., for analysis of signal coverage in a city) \citep{choi2023withray,de2005faspro}.
Only a few select works are cited.
Due to their importance, these applications have garnered significant attention, and better solutions would be of real interest.

SBR and UTD-based ray tracing methods are perfectly local, and some may be complete, but as discussed, such methods can never be linear.
That is, rays are always \emph{mutually interfering}, and the interference term in \cref{bilinear_superposition} must always be considered.
As very many rays need to be traced, attempts to accelerate SBR methods have taken a research trajectory not dissimilar to early computer graphics work: 
Employing multi-resolution grids \cite{Suk2001} and spatial-subdivision data structures to accelerate ray-facet intersections \cite{Jin2006,Tao2008}; accelerating on GPUs \cite{YuboTao2010,gao2015}; and, sampling the initial ray directions using Halton sequences \cite{Key2018}.
Nevertheless, as linearity is unavoidably lost, this research greatly trails computer graphics in its ability to apply sophisticated sampling techniques, like importance sampling light-matter interactions or path guiding.
Scene complexity remains a very limiting factor.

\paragraph{Holography}

Some limited forms of wave-optical light transport have been used for computer-generated holography (CGH).
These algorithms share similar problems to the methods described above.
\citet{magallon2021slm} propose to trace ray tubes backwards and accumulate the coherent contributions from all traced paths.
\citet{Blinder:21} also consider the challenges of efficient backward wave-optical path tracing, and propose a method which discretizes and integrates a diffraction integral.
Being perfectly-local but non-linear formalisms, both suffer from the issues of non-linearity.
An accurate backward model of wave-optical light transport is important for CGH.
Our formalism could be applied to CGH in the future, though it is beyond the scope of this paper.

\paragraph{Optical speckle}

Tangentially related is the rendering of speckle or the integration of speckle statistics \cite{Bar_Alterman_Gkioulekas_Levin_2019,Bar_Gkioulekas_Levin_2020}.
Speckle are patterns produced by coherent contributions from many scatterers.
However, not all such contributions give rise to \emph{observable} speckle.
Observable optical phenomena are always \emph{integrated} effects: the integration is over the spatial and angular extent of the observer (which, by the uncertainty principle, must be positive), and over the period of observation.
Point-sampling interference effects (as perfectly-local formalisms most often do) suffer from aliasing.
This is discussed in greater detail in \cref{section_gr_analysis}. 
We show in \cref{fig_twin_slit} that as we integrate over the extent of an observer, scatterers that are sufficiently far apart do not contribute to observable interference effects, even in the presence of perfectly-coherent illumination.

Typical speckle integration algorithms (and forward light transport formalisms) have no means to establish whether a pair of scatterers do indeed contribute to observable speckle, and hence can be wasteful.
In our formalism, generalized rays quantify exactly the spatial extent in a scene over which observable interference may indeed arise.
Therefore, our formalism could be used in future work to speed up such speckle rendering, as well as support partially-coherent speckle.
A sample-solve method could also be used to integrate speckle statistics.
This is out of scope of this paper.

\paragraph{Material appearance reproduction}

Also related is work that aims to reproduce the appearance of some diffractive materials.
This includes the rendering of iridescent and pearlescent materials \cite{Guillen:2020:Pearlescence}; diffractive scratches \cite{Velinov2018scratches,Werner2017Scratch}; statistical surface profiles \cite{Krywonos2006,Holzschuch2017,Steinberg_speckle_2021} or diffractive surfaces with explicit microgeometry \cite{Yu_Xia_Walter_Michielssen_Marschner_2023,Falster2020}; and, thin-film interference at a soap bubbles \cite{Huang_Iseringhausen_Kneiphof_Qu_Jiang_Hullin_2020} or due to a dielectric layer over a conductor \cite{Belcour:17,Kneiphof2019}.
Synthesis of BSDFs that account for wave interference was discussed by \citet{Toisoul2017diffractions}.

The cited work does not deal with light transport in a scene, but only considers the result of an interaction with a material.
Assumptions are made regarding the structure of the simulated fields: these are often restricted to plane waves, which can only be used for light transport under an asymptotic, far-field approximation; or, Gaussian beams, but a decomposition into mutually-incoherent Gaussian beams in a forward model is a crude approximation (and always suffers from the sampling problem).
In this work we are not concerned with appearance reproduction, or the simulation of some particular wave-optical interaction at the material level. 
Instead, we focus on deriving a light transport formalism that does not suffer from the sampling problem.


\section{Background: Theoretical Foundations} \label{section_theory}

We briefly introduce the Wigner picture of wave optics, which is a useful setting for the discussion of the dynamics of light. 
See our supplemental material or \citet{torre2005linear,testorf2010phase} for additional information.
We also formalize the process of photoelectric detection \cite{ou1995probability,mandel1995optical}. 
Our interest in photoelectric detection is motivated by how our backward light transport formalism (introduced in \cref{section_backward_light_transport}) operates:
In contrast to forward models of light transport, which propagate the emission distributions of light forward in time, we propagate the \emph{detection states} of a detector under time-reversed dynamics.
This is key in decoupling the light transport simulation from the properties of light, thereby solving the sampling problem.

\subsection{Phase-Space Optics}

The basic descriptor of light under wave optics is the \emph{wave function}, denoted $\psi(\va{r} \argsep t)$, where $\va{r}$ is position and $t$ is time.
The wave function quantifies the spatial excitations of a component of the associated electric field.
Henceforth, we will fix time and, for brevity, drop $t$ from the argument lists.
In his seminal work in 1932, Wigner formulated the \emph{Wigner distribution function} (WDF) \cite{Wigner1932}, which describes a \emph{position-momentum} distribution of the signal $\psi$:
\begin{align}
    \wvd\qty(\va{r},\va{k}) 
    &\triangleq
        \tfrac{1}{\qty(2\mpi)^3}
        \int
            \dd{\va{r}^\prime} 
            \psi^\star\qty(\va{r} - \tfrac{1}{2} \va{r}^\prime)
            \psi\qty(\va{r} + \tfrac{1}{2} \va{r}^\prime)
            \ee^{-\ii \va{r}^\prime\cdot\va{k}}
    \label{def_WDF}
    ~,
\end{align}
where $\star$ denotes complex conjugation, $\va{r}$ is a spatial position, and $\va{k}$ is the \emph{wavevector}, which quantifies light's temporal frequency and direction of propagation.
$\abs*{\va{k}}=\eta\tfrac{2\mpi}{\lambda}$ is the \emph{wavenumber}, with $\eta$ being the refractive index of the medium in which light propagates and $\lambda$ is the light's wavelength.
Up to a phase term, the wave function may always be recovered from the WDF, hence the WDF provides a complete description of light.

The domain of the WDF is a space of both positions and wavevectors, referred to as \emph{phase space}. 
This phase-space treatment of wave optics admits stark similarities to the ray-optical depiction of light: $\wvd(\va{r},\va{k})$ is the energy density of a ``light particle'' at position $\va{r}$ with direction of propagation proportional to $\va{k}$.
$\wvd(\va{r},\va{k})$ also propagates like the classical ray in free space and under some other interactions.
Furthermore, it is well known that the WDF fulfils most of the postulates expected of a classical (ray optical) phase-space density function. 
The WDF does depart from the ray-optical picture in a couple of fundamental aspects:
\begin{enumerate}
    \item   \textbf{Bilinearity}:
            The WDF is a bilinear distribution, hence a linear superposition of WDFs is not the WDF of a superposition of waves (see \cref{wdf_bilinear}).
    \item   \textbf{Negativity}:
            The WDF takes somewhat counter-intuitive negative values, restricting its interpretation as an energy density.
            These negative values are not an exception, as a matter of fact, non-negative WDFs arise only with one specific class of wave functions \cite{torre2005linear}.
\end{enumerate}

As discussed, the properties of the WDF have motivated other work to use the WDF to perform ray-like ``point queries'' of light behaviour for wave-optical light transport. 
However, either locality is entirely lost when interaction with matter considered (viz. \cref{wdf_convolution_kernel}), because the kernel $K$ must account for the entire scene; or, if we restrict $K$ to a subset of the scene, then WDFs that arise from interactions with different such kernels must superpose bilinearly.
No formalism can be simultaneously perfectly-local and linear.
In addition, as a consequence of the WDF not being non-negative, it is a highly-oscillatory function, as shown in \cref{section_rendering_wave_optics} in our supplemental material.
This means that integration of the WDF with point samples heavily suffers from aliasing.

We are not interested in working with potentially-arbitrary emitted WDFs, as done by other Wigner-based formalisms.
Instead, we work in phase space to study the dynamics of \emph{detection states}, introduced next, and show that these states admit far nicer properties than a general WDF.

\paragraph{Gaussian distributions}

A phase-space construct central to our discussion is the (symplectic) \emph{Gaussian WDF} 
\cite{testorf2010phase}:
\begin{gtcb}{cyan!4!white}{black!90!gray}{blue!20!cyan}{\vphantom{A}\smash{Phase-space Gaussian}}%
\begin{align}%
    g_{\beta,\rho}\qty(\va{r},\va{k} \argsep \va{r}_0,\va{k}_0)
        \triangleq&
            \tfrac{1}{\mpi^3}
            \exp[
                 -\tfrac{1+\rho^2}{\beta^2} \abs{\vphantom{T^a}\smash{\va{r}-\va{r}_0}}^2
                 -\beta^2 \abs{\vphantom{T^a}\smash{\va{k}-\va{k}_0}}^2
            ]
    \nonumber \\ 
        &\quad\times
            \exp[2\rho 
                 \abs{\vphantom{T^a}\smash{\va{r}-\va{r}_0}}
                 \abs{\vphantom{T^a}\smash{\va{k}-\va{k}_0}}
            ]
    \label{symplectic_Gaussian}
    ~.
\end{align}%
\end{gtcb}%
\vspace*{-1mm}
\noindent
As will be discussed in \cref{section_photoelectric_measurement}, the (non-negative) values measured by our detectors can always be written as, and only as, a convolution of such phase-space Gaussians with an arbitrary WDF.
Therefore, we will use these Gaussians as our weakly-local light transport primitives.

The parameterization of the Gaussian is chosen such that we may understand $\beta>0$ as the initial spatial variance of the WDF $g$, and the correlation parameter $\rho\geq 0$ is related to propagation distance.
As we will see later, propagation of light induces phase-space correlation between the spatial and wavevector variables, i.e. propagation increases $\rho$.
In general, $\beta,\rho$ can be positive-definite matrices, describing anisotropy (see our supplemental material).
For simplicity, here we assume scalar parameters.
$g$ is normalized such that $\int \smash{\dd{\va{r}} \dd{\va{k}} g_{\beta,\rho}} = 1$.

Phase-space Gaussians can also be understood as the most compact physically-realizable construct:
note that the product of the spatial and wavevector variances of the Gaussian above is $\tfrac{1}{4}(1+\beta^4\rho^2)$, which fulfils the \emph{uncertainty relation} \cite[Chapter 4]{mandel1995optical}, viz. $\sigma_r^2\sigma_k^2=\quarter$, if and only if $\rho=0$.

\subsection{Measurement of the WDF \hardsec} \label{section_photoelectric_measurement}

In sharp contrast to classical physics, we may not measure light without disturbing it: 
under wave optics, any measurement apparatus deduces information about light by interacting with it.
The action of a detector on light can be quantified by the detector's WDF, and the intensity observed by the detector is \cite{dragoman2005phase}
\begin{align}
    I =&
        \int \dd{\va{r}^\prime} \dd{\va{k}^\prime} 
            \wvd\qty(\va{r}^\prime,\va{k}^\prime)
            \wvd[d]\qty(\va{r}^\prime,\va{k}^\prime)
    \label{wdf_measurement}
    ~,
\end{align}
where $\wvd$, $\wvd[d]$ are the WDFs of light and the detector, respectively.
Even though the WDF takes negative values, the result of a measurement is always non-negative, i.e. $I\geq 0$, which formally follows directly from the Moyal formula \cite[Chapter 6.3.7]{torre2005linear}.

We assume our detectors are \emph{classical photoelectric} detectors.
Photons impinging upon a photoelectric detector ionize free electrons, producing a current that is amplified and measured.
All detectors of interest (e.g., a photoreceptor in the eye, a pixel in a CMOS array in a camera, or a RADAR antenna) operate in this fashion.
Classicality implies that quantum effects are ignored: we assume that the photon flux is sufficiently high that electric pulses from single photons overlap and form a continuous electric current. 

As photoelectric detection works by absorption of photons, it is the \emph{photon annihilation} and \emph{photon number} operators that are the observables \cite[Chapter 11]{mandel1995optical}.
The only eigenstates of these operators are the (squeezed) \emph{coherent states} \cite[Chapter 11, 21]{mandel1995optical}, whose phase-space representation takes the form of uncorrelated (minimum-uncertainty) Gaussians, i.e. the Gaussians defined in \cref{symplectic_Gaussian} with $\rho=0$, viz.
\begin{align}%
    \hat{g}_{\beta}\qty(\va{r},\va{k} \argsep \va{r}_0,\va{k}_0)
        \triangleq&
            \tfrac{1}{\mpi^3}
            \ee^{-\frac{1}{\beta^2} \abs{\va{r}-\va{r}_0}^2}
            \ee^{-\beta^2 \abs*{\va{k}-\va{k}_0}^2}
    \label{coherent_state}
    ~.
\end{align}%
As with the general Gaussian $g$, the parameter $\beta>0$ above quantifies the trade-off between the spatial and wavevector variances.

Let a detector be described by a distribution $\mathcal{D}$ of detection elements, each quantified by the parameters:
\begin{enuminline}
    \item $\va{k}_0$: mean wavevector of detection (quantifies the mean wavelength and direction of propagation of detected light);
    \item $\beta$: spatial extent; and, 
    \item $\alpha$: detection efficiency (quantum efficiency).
\end{enuminline}
The WDF $\wvd[d]$ of the detector quantifies which light states are detectable \cite{dragoman2005phase}.
As the states detectable by a classical photoelectric detector are the coherent states, the detector's WDF may be written as
\begin{align}
    \wvd[d]\qty(\va{r},\va{k}) 
        =&
            \int_{\mathcal{D}} \dd{\mu_\mathcal{D}}
                \alpha
                \hat{g}_{\beta}\qty(\va{r},\va{k} \argsep \va{r}_0,\va{k}_0)
    \label{wdf_coherent_states}
\end{align}
where $\mu_\mathcal{D}$ is the measure over $\mathcal{D}$.
In the simplest case, $\va{k}_0,\alpha,\beta$ are constant, and $\mathcal{D}$ quantifies the spatial region occupied by the detector.
Then, the integration is over the spatial extent, with $\va{r}_0$ being the integration variable.
In general, $\mathcal{D}$ may quantify spatially-varying detection properties.

Denote the \emph{Husimi Q distribution} of a WDF $\wvd$ as the convolution of the WDF with a phase-space Gaussian:
\begin{align}
    \husimiQ\qty(\va{r},\va{k},\beta)
        \triangleq&
            \int
                \dd{\va{r}^\prime} \dd{\va{k}^\prime} 
                \wvd\qty(\va{r}^\prime,\va{k}^\prime)
                \hat{g}_{\beta}\qty(\va{r}^\prime,\va{k}^\prime \argsep \va{r}_0,\va{k}_0)
    \label{def_husimiQ}
    ~.
\end{align}
Substitute \cref{wdf_coherent_states,def_husimiQ} into \cref{wdf_measurement} and formally interchange the orders of integration:
\begin{align}
    I =&
        \int \dd{\va{r}^\prime} \dd{\va{k}^\prime} 
            \wvd\qty(\va{r}^\prime,\va{k}^\prime)
            \int_{\mathcal{D}} \dd{\mu_\mathcal{D}}
                \alpha
                \hat{g}_{\beta}\qty(\va{r}^\prime,\va{k}^\prime \argsep \va{r}_0,\va{k}_0)
    \nonumber \\
    =&
        \int_{\mathcal{D}} \dd{\mu_\mathcal{D}}
            \alpha
            \husimiQ\qty(\va{r}_0,\va{k}_0,\beta)
    \label{husimiQ_measurement}
    ~.
\end{align}
The above implies that a classical photoelectric detector directly measures the Husimi Q distribution, and not the WDF.
The ``smoothing'' operation of the WDF over a minimum-uncertainty Gaussian (as in \cref{def_husimiQ}) is a sufficient and necessary condition to produce a bandwidth-limited, non-negative distribution \cite{soto1983wigner}, and it is the smoothed distribution that classical photoelectric detectors observe \cite{dragoman2004phase,Leonhardt1997-xz}.


\section{Theory of Backward Wave-Optical Light Transport} \label{section_backward_light_transport}

Note the symmetry in the measurement formula, \cref{wdf_measurement}: we may understand it as the detector acting upon the incident distribution of light; or, equivalently, as light acting upon detection states.
A forward model works by sourcing a light distribution $\wvd$ from light sources, simulating its interaction with the scene, and finally integrating it over the detector's distribution.
On the other hand, our backward model that we present in this Section evolves the detector distribution $\wvd[d]$ under time-reversed dynamics, and then integrates it over the sourcing distribution.
In contrast to the sourced WDF, the WDF of the photoelectric detector $\wvd[d]$ (\cref{wdf_coherent_states}) admits much nicer analytic properties.

We term the detection states' wave function as the \emph{generalized ray}.
We formally define the generalized ray and discuss their sourcing from an arbitrary photoelectric detector $\mathcal{D}$ in \cref{section_generalized_rays}.
In \cref{section_light_transport_gr}, we will formalize the workings of our backward light transport formalism with generalized rays.
In \cref{section_gr_analysis} we analyze this formalism and show that it is weakly-local, linear and complete. 
In \cref{section_interaction_kernels_examples} the interaction kernels will be discussed further, and we will show how to derive such kernels.
In \cref{section_rendering} we build upon the formalism developed here, and present a practical, interactive wave-optical renderer.

\paragraph{Time-reversed dynamics}

Consider the WDF $\wvd[s]$ of light that is sourced from one or more light sources.
Let the light quantified by $\wvd[s]$ propagate around a scene, interact with that scene, and finally impinge upon a classical photoelectric detector, with detection WDF $\wvd[d]$, as in \cref{wdf_coherent_states}.
Let the kernel for the entire process be denoted $K$ (as defined in \cref{WDF_kernel_def}), and we denote the action of such kernel on a WDF (as in \cref{wdf_convolution_kernel}) via the \emph{interaction operator}:
\begin{align}%
    \mathpzc{K}\qty{\wvd}\qty(\va{r}_o,\va{k}_o) 
    \triangleq&
        \int 
            \dd{\va{r}_i} \dd{\va{k}_i} 
            K\qty(\va{r}_o,\va{r}_i,\va{k}_o,\va{k}_i) 
            \wvd\qty(\va{r}_i,\va{k}_i)
    \label{interaction_operator}
    ~.
\end{align}%
Denote the WDF of the light that impinges upon the detector, after interaction with the scene, as $\wvd[o] = \mathpzc{K}\{\wvd[s]\}$. 
Apply the measurement formula, \cref{husimiQ_measurement}, to $\wvd[o]$, yielding an expression for the result of a measurement of light after interaction:
\begin{align}
    I =&
        \int_{\mathcal{D}} \dd{\mu_\mathcal{D}}
            \alpha
            \husimiQ[o]\qty(\va{r}_0,\va{k}_0,\beta)
    \nonumber \\
    =&
            \int_{\mathcal{D}} \dd{\mu_\mathcal{D}}
                \alpha\!
                \int\! \dd{\va{r}^\prime} \dd{\va{k}^\prime} 
                    \hat{g}_{\beta}\qty(\va{r}^\prime,\va{k}^\prime \argsep \va{r}_0,\va{k}_0)
                    \mathpzc{K}\qty{\wvd[s]}\qty(\va{r}^\prime,\va{k}^\prime)
    \label{gen_rays_measurement_wvd1}
    ~,
\end{align} 
where $\husimiQ[o]$ is the Husimi Q distribution (\cref{def_husimiQ}) of $\wvd[o]$.

Time-reversal induces wavevector reversal, viz. $\va{k}\to-\va{k}$, and phase conjugation \cite[Chapter 2.3]{Geru_2018}.
Hence, a system's optical response function $h$ fulfils $h(\va{r}_o,\va{r}_i)=h^\star(\va{r}_i,\va{r}_o)$.
By applying this relation to \cref{WDF_kernel_def}, 
we define the \emph{time-reversed diffraction kernel}:
\begin{align}
    K^{\timereversal}\qty(\va{r}_o,\va{r}_i,\va{k}_o,\va{k}_i)
        \triangleq& 
            K^\star\qty(\va{r}_i,\va{r}_o,\va{k}_i,\va{k}_o)
    \label{def_inverse_kernel}
    ~.
\end{align}
The time-reversed interaction operator $\mathpzc{K}^{\timereversal}$ is then defined to act upon a WDF via $K^{\timereversal}$, as in \cref{interaction_operator}.
Using \cref{interaction_operator,WDF_kernel_def}, we may observe that $\mathpzc{K}^{-1}\equiv\mathpzc{K}^{\timereversal}$, for lossless kernels.
Formally interchange the integration order in \cref{gen_rays_measurement_wvd1} between the integral over the primed variables and the integral over the kernel $K$, and apply the time-reversed interaction operator, yielding
\begin{gtcb}{gray!5!white}{gray!3!white}{gray!22!black}{\vphantom{A}\smash{Backward light transport formalism}}%
\begin{align}%
    \!\!\!
    I =& 
        \int_{\mathcal{D}} \dd{\mu_\mathcal{D}}
            \alpha
            \int \dd{\va{r}^\prime} \dd{\va{k}^\prime} 
                \wvd[s]\qty(\va{r}^\prime,\va{k}^\prime)
                \mathpzc{K}^{\timereversal}\qty{
                    \hat{g}_{\beta}
                }\qty(\va{r}^\prime,\va{k}^\prime)
    \label{backward_light_transport_formalism}
    ~.
\end{align}%
\end{gtcb}%
\vspace*{-1mm}
\noindent

Subject to the physics of photoelectric detection (\cref{section_photoelectric_measurement}), all the derivations above are exact.
\cref{gen_rays_measurement_wvd1} quantifies the \emph{forward} model of light transport: a distribution of light sourced from light sources, $\wvd[0]$, is evolved forward in time, and then measured by a detector.
On the other hand, \cref{backward_light_transport_formalism} quantifies the \emph{backward} model: each detector state of a photoelectric detector is evolved backward, under the time-reversed wave-optical dynamics (as quantified by \cref{def_inverse_kernel}), and then the overlap of this state with the sourcing distributions are computed.
Both yield exactly the same answer.
However, in contrast to the potentially arbitrary WDF $\wvd[0]$, the artefacts of photoelectric detection---the states $\hat{g}$---are much ``better behaved'' functions: they are simple Gaussians, bandwidth limited and always non-negative.

\subsection{Generalized Rays \hardsec} \label{section_generalized_rays}

By applying the inverse WDF transform to $g$ \cite[Chapter 6.3.5]{torre2005linear}, we derive the wave function of the phase-space Gaussian, which we term the \emph{generalized ray}:
\begin{gtcb}{blue!4!white}{blue!50!black}{cyan!40!white}{\vphantom{A}\smash{Generalized ray (wave function)}}%
\begin{align}%
    \!\!
    \psi_{\beta,\rho}\qty(\va{r} \argsep \va{r}_0,\va{k}_0) \triangleq&
        \qty(\tfrac{1}{\mpi\beta^2})^{\threequarters}
        \ee^{\ii \va{k}_0 \cdot \qty(\va{r}-\va{r}_0)}
        \ee^{-\smash{\tfrac{1}{2\beta^2}}\qty(1 - \ii\rho) \abs{\va{r}-\va{r}_0}^2}
    \label{def_gen_ray}
    ~.
\end{align}%
\end{gtcb}%
\vspace*{-1mm}
\noindent
A general expression with anisotropic $\beta,\rho$ is available in our supplemental material.
The generalized ray is defined with respect to the mean spatial position $\va{r}_0$, mean wavevector $\va{k}_0$, and parameters $\beta,\rho$.
The reader may verify via \cref{def_WDF} that the phase-space representation of the generalized ray (its WDF) is $g_{\beta,\rho}(\va{r},\va{k} \argsep \va{r}_0,\va{k}_0)$.

Initially, i.e. given $\rho=0$, a generalized ray is one of the detection states of the detector $\mathcal{D}$, and its phase-space representation is a coherent state $\hat{g}_\beta$ that minimizes the product of its space and wavevector variances.
Thus, it is the \emph{most local} construct permissible under wave optics \cite{torre2005linear}.
As it name suggests, the generalized ray can be understood as \emph{the closest analogue to the classical ray under wave optics}.
Once it is propagated away from the detector ($\rho>0$), its WDF becomes the correlated Gaussian $g$, as in \cref{symplectic_Gaussian}.

\paragraph{Sourcing}

Let $\wvd[s]$ be a sourcing WDF, as before.
Under the backward model of light transport (\cref{backward_light_transport_formalism}), the result of a measurement
at a detector element $\{\va{r}_0, \va{k}_0, \beta, \alpha\} \in \mathcal{D}$ can be written as 
\begin{align}
    I\qty(\va{r}_0,\va{k}_0,\beta)
    \triangleq&
        \int \dd{\va{r}^\prime} \dd{\va{k}^\prime} 
            \wvd[s]\qty(\va{r}^\prime,\va{k}^\prime)
            \mathpzc{K}^{\timereversal}\qty{\hat{g}_{\beta}}\qty(\va{r}^\prime,\va{k}^\prime)
    \label{measurement_at_detector_element_backward}
    ~.
\end{align}
The total intensity integrated over the entire detector is then 
\begin{align}
    I =& 
        \int_{\mathcal{D}} \dd{\mu_\mathcal{D}}
            \alpha
            I\qty(\va{r}_0,\va{k}_0,\beta)
    \label{total_intensity_D}
    ~.
\end{align}

By Monte-Carlo integrating the integral over the detector distribution $\mathcal{D}$ in \cref{total_intensity_D}, we \emph{source} a generalized ray from the detector: 
\begin{gtcb}{violet!3!white}{violet!50!black}{violet!65!cyan!30!white}{\vphantom{A}\smash{Sourcing}}%
\begin{align}%
    \!\!
    I \approx& 
        \frac{1}{N} 
        \sum_{n=1}^{N} 
            \frac{
                \alpha^{(n)}
            }{p^{(n)}}
            I\qty(\va{r}_0^{(n)},\va{k}_0^{(n)},\beta^{(n)})
    \nonumber \\
    =&
        \frac{1}{N} 
        \sum_{n=1}^{N} 
            \frac{
                \alpha^{(n)}
            }{p^{(n)}}
            \!\!
            \int\! 
                \dd{\va{r}^\prime} \dd{\va{k}^\prime} 
                \wvd[s]\qty(\va{r}^\prime,\va{k}^\prime)
                \mathpzc{K}^{\timereversal}\qty{g^{(n)}}\qty(\va{r}^\prime,\va{k}^\prime)
    \label{MC_sourcing}
\end{align}%
\end{gtcb}%
\vspace*{-1mm}
\noindent
(normalization constants accounted for by $\alpha$), with $p^{(n)}$ being the sampling probabilities.
Each sampled 
{\small $\{\va{r}{}_0^{(n)},\!\va{k}{}_0^{(n)},\!\beta^{(n)},\!\alpha^{(n)}\}\!\in\!\mathcal{D}$} 
is a detector state that quantifies a sourced generalized ray, whose WDF is denoted as
\begin{align}
    g^{(n)}\qty(\va{r},\va{k}) \triangleq& \hat{g}_{\beta^{(n)}}\qty(\va{r}, \va{k} \argsep \va{r}{}_0^{(n)},\va{k}{}_0^{(n)})
    \label{sampled_generalized_ray}
\end{align}
(recall that detection state $\hat{g}_\beta$ is simply $g_{\beta,\rho}$ with $\rho=0$).
A sourced generalized ray is defined by its spatial position on the detector {\small$\va{r}{}_0^{(n)}$}, the wavevector {\small$\va{k}{}_0^{(n)}$} that quantifies wavelength and direction of propagation of the generalized ray from the detector, and the variance $\beta^{(n)}$ that quantifies its spatial extent.
At sourcing, we set $\rho^{(n)}=0$.
At the detector, $\alpha$ is the detection efficiency, however once a generalized ray is sourced $\alpha^{(n)}$ should be understood as the intensity carried by that generalized ray.

\subsection{Light Transport with Generalized Rays \hardsec} \label{section_light_transport_gr}

For each generalized ray sampled from a detector, viz. \cref{MC_sourcing}, we run a light transport simulation, as formalised by \cref{backward_light_transport_formalism}.
That simulation is dictated by the time-reversed (total) interaction operator $\mathpzc{K}^{\timereversal}$.
To compute the contributions of the detection element to the total intensity observed by the detector, we finally integrate the generalized ray over the light sources' sourcing distributions $\wvd[s]$.

Often it is more practical to formulate the interaction operator $\mathpzc{K}$, that quantifies the total action on light of the entire scene, as a sequence of interactions, viz. 
$\mathpzc{K} \equiv \mathpzc{K}_1 \composition \mathpzc{K}_2 \composition \ldots \composition \mathpzc{K}_M$.
A single interaction $\mathpzc{K}_{\ j}$ might now describe free-space propagation, or a light-matter interaction.
The definition of the composed interaction operator $\mathpzc{K}_{\ l} \composition \mathpzc{K}_{\ m}$ only depends on the respective interaction kernels $K_l,K_m$, and follows immediately from \cref{wdf_convolution_kernel} (see \citet[Chapter 8.3.2]{torre2005linear}).
The time-reversed interaction operator trivially becomes
$
    \mathpzc{K}^{\timereversal} \equiv
        \mathpzc{K}_M^{\timereversal} \composition \mathpzc{K}_{M-1}^{\timereversal} \composition \ldots \composition \mathpzc{K}_{\ 1}^{\timereversal}
$.

We will now recursively apply each kernel $\mathpzc{K}^{\timereversal}_m$ to a generalized ray, in order to Monte Carlo integrate the light transport through the scene.
This process is similar to classical light transport:
A generalized ray undergoes interactions with the scene at discretized steps.
Because a generalized ray is weakly-local (as will be discussed in \cref{section_gr_analysis}), each interaction is confined to a small region in space.
A sequence of interactions gives rise to a \emph{path}, enumerated via the indices of the generalized rays chosen at each interaction, starting with $n$, the index of the sourced generalized ray from \cref{MC_sourcing}.
Let a path of depth $m-1$ be denoted by a tuple $\vb{\pi}=(n_1,\ldots,n_m)$, where the indices $n_i$ denote sampled generalized rays at each interaction. 
We define the addition operation of a path with an index as: $\vb{\pi}+n_{m+1}=(n_1,\ldots,n_m,n_{m+1})$, for some index $n_{m+1}$.

We denote a generalized ray's phase-space representation as the Gaussian WDF $g^{\vb{\pi}}$, for an arbitrary path $\vb{\pi}$, as
\begin{align}
    &g^{\vb{\pi}}\qty(\va{r},\va{k}) 
    \triangleq 
        g_{\beta^{\vb{\pi}},\rho^{\vb{\pi}}}
        \qty(\va{r}, \va{k} \argsep \va{r}_0^{\vb{\pi}}, \va{k}{}_0^{\vb{\pi}})
    \label{light_transport_husimiQ_generalized_ray}
    ~,
\end{align}
each defined w.r.t. the values 
{\small$\beta^{\vb{\pi}},\rho^{\vb{\pi}},\va{r}{}_0^{\vb{\pi}},\va{k}{}_0^{\vb{\pi}}$}.
Note that \cref{light_transport_husimiQ_generalized_ray} generalizes \cref{sampled_generalized_ray}.

\paragraph{Recursive light transport}

Let $\vb{\pi}$ be some path of depth $m-1\geq 0$, i.e. $\abs*{\vb{\pi}}=m$, and $g^{\vb{\pi}}$ the generalized ray sampled using that path.
For the initial case $m=1$, $g^{\vb{\pi}}=g^{(n)}$ is a sampled sourced generalized ray from the detector, as in \cref{MC_sourcing,sampled_generalized_ray}.
Denote the WDF that arises after interaction with $\mathpzc{K}_{\ m}^{\timereversal}$:
\begin{align}
    \!\!\!
    \husimiQ[m]^{\vb{\pi}}
    \triangleq
        \mathpzc{K}_{\ m}^{\timereversal}\qty{g^{\vb{\pi}}}
    =
    \!\!
        \int 
            \dd{\va{r}_i} \dd{\va{k}_i} 
            K^{\timereversal}_m\qty(\va{r},\va{r}_i,\va{k},\va{k}_i) 
            g^{\vb{\pi}}\qty(\va{r}_i,\va{k}_i)
    \label{light_transport_Q1}
    ~.
\end{align}
As interaction kernels fulfil all the properties of a WDF \cite[Chapter 1.6]{testorf2010phase}, the integration above
is a phase-space convolution of a WDF with a Gaussian, hence $\husimiQ[m]^{\vb{\pi}}$ is indeed a Husimi Q distribution (as defined in \cref{def_husimiQ}).
Note, the above holds only because our generalized ray is a phase-space Gaussian (a WDF is a non-negative husimi Q distribution if, and only if, it can be written as a convolution of an arbitrary WDF with a Gaussian).

Gaussian functions form an overcomplete functional basis \cite{Rudin1990-ak}.
Therefore, a Husimi Q distribution, being a non-negative, bandwidth-limited distribution (a consequence of the convolution), may always be written as a finite sum of Gaussians to arbitrarily high accuracy, viz.
\begin{align}
    \husimiQ[m]^{\vb{\pi}}\qty(\va{r},\va{k}) 
    \approx&
        \frac{1}{N_m}
        \sum_{\vphantom{a}\smash{n_m=1}}^{\vphantom{a}\smash{N_m}}
            \alpha^{\vb{\pi}+(n_m)}
            g^{\vb{\pi}+(n_m)}\qty(\va{r},\va{k})
    \label{MC_light_transport_Q1}
    ~,
\end{align}
where $\smash{\alpha^{\vb{\pi}+(n_m)}}>0$ are the intensities of each Gaussian above (positiveness is a consequence of the non-negativity of a Husimi Q distribution).
Thereby, we have sampled $N_m$ generalized rays with paths $\vb{\pi}+(n_m)$ of depth $m$, arising from the action of the operator $\mathpzc{K}_{\ m}$ on the generalized ray $g^{\vb{\pi}}$.

Continue this process recursively, giving rise to sampled generalized rays $g^{\vb{\pi}_j}$, with $\vb{\pi}_j$ being the sampled paths, and each generalized ray is defined by the tuple 
{\small $\{\va{r}{}_0^{\vb{\pi}_j}, \va{k}{}_0^{\vb{\pi}_j}, \beta^{\vb{\pi}_j}, \rho^{\vb{\pi}_j}\}$} and {\small $\alpha^{\vb{\pi}_j}$}.
The sampled path depth is $\abs*{\vb{\pi}_j}-1$:
For $\abs*{\vb{\pi}_j}=1$ these generalized rays are sourced from the detector $\mathcal{D}$, while for $\abs*{\vb{\pi}_j}>1$ these generalized rays are sourced from a distribution that arises after interaction with 
$\mathpzc{K}_{\ \abs*{\vb{\pi}_j}-1}^{\timereversal} \composition \ldots \composition \mathpzc{K}_{\ 1}^{\timereversal}$.

\paragraph{The rendering equation}

The recursive \emph{rendering equation for backward wave-optical light transport}, that drives the process above, takes the familiar form of a Fredholm integral equation:
\begin{gtcb}{orange!5!white}{purple!60!black}{purple!70!blue!30!white}{\vphantom{A}\smash{Rendering equation}}%
\begin{align}%
    \!\!
    \eval{\husimiQ[o]}_{\va{r}_0}
    =&
        \!\int\!
            \dd{\va{k}_0} \dd{\beta} \dd{\rho}
            \mathpzc{K}_{\ r}^{\timereversal}\qty{
                g_{\beta,\rho}\qty(\va{r}^\prime,\va{k}^\prime \argsep \va{r}_0, \va{k}_0)
            }
    \label{rendering_equation}
    ~,
\end{align}%
\end{gtcb}%
\vspace*{-1mm}
\noindent
where $\mathpzc{K}_{\ r}^{\timereversal}$ is some interaction operator (acts w.r.t. the primed variables), $g$ is an incident generalized ray and the integration is over all possible generalized rays' wavevectors $\va{k}_0$ and parameters $\beta,\rho$.
In contrast to the classical rendering equation that acts upon scalars, \cref{rendering_equation} acts upon functions, all generalized rays that are incident to mean position $\va{r}_0$, and its result is a function as well, the distribution $\husimiQ[o]$.
Outgoing generalized rays are then sampled from the resulting light distribution $\husimiQ[o]$, in an identical manner to \cref{MC_light_transport_Q1}.
There is no emission term: interaction with light sources is handled by the measurement operator, introduced next.
The interaction operator will be discussed further in \cref{section_interaction_kernels_examples}.

\paragraph{Measurement}

The recursive light transport process above solves for the total light transport, viz. the term $\mathpzc{K}^{\timereversal}\{g^{(n)}\}$ in \cref{MC_sourcing}.
The observed intensity is computed by integrating each generalized ray over the sourcing distribution $\wvd[s]$, quantified by the following \emph{measurement operator}:
\begin{gtcb}{red!5!white}{red!40!black}{red!40!orange!30!white}{\vphantom{A}\smash{Measurement}}%
\begin{align}
    \!
    \mathfrak{I}\qty{g} 
    \triangleq&
        \int\! 
            \dd{\va{r}^\prime} \dd{\va{k}^\prime} 
            \wvd[s]\qty(\va{r}^\prime,\va{k}^\prime)
            g_{\beta,\rho}\qty(\va{r}^\prime,\va{k}^\prime \argsep \va{r}_0, \va{k}_0)
    \nonumber \\
    =&
        \tfrac{1}{\qty(2\mpi)^3}
        \abs{
        \int\! \dd{\va{r}^\prime}
            \psi_s\qty(\va{r}^\prime)
            \psi_{\beta,\rho}^\star\qty(\va{r}^\prime \argsep \va{r}_0, \va{k}_0)
        }^2
    \label{measurement_operator}
    ~,
\end{align}%
\end{gtcb}%
\vspace*{-1mm}
\noindent
where we apply the Moyal formula \cite[Chapter 6.3.7]{torre2005linear} in-order to reformulate the phase space integration using Wigner distributions into integration in position space using wave functions,
and with $\psi_s$ being the sourced wave function.
In practice, the integration above happens once a generalized ray is incident upon a light source, and under the support of that light source's WDF.

In \cref{section_rendering,alg_rendering} we will develop a practical light transport algorithm, building upon the derivations in this Section.


\section{Analysis} \label{section_gr_analysis}

In \cref{section_backward_light_transport} we have derived our primary contributions: the theory of backward wave-optical light transport with generalized rays.
We now analyse and validate these results.

\paragraph{Linearity}

All formulations above are linear: 
\begin{enuminline}
    \item Sourcing equations, \cref{total_intensity_D,measurement_at_detector_element_backward}, describe a linear (incoherent) superposition of the measured intensities of each detector element;
    \item the rendering equation, \cref{rendering_equation}, is linear; and,
    \item measurement of a generalized ray, i.e. integration over the sourcing distribution (\cref{measurement_operator}), as well as accumulation of these intensities are also linear operations. 
\end{enuminline}
Generalized rays always carry positive intensities.
We discuss the linearity of generalized ray further, from the perspective of Shannon sampling, in \cref{subsection_linearity_of_generalized_rays}.

\paragraph{Locality}

We need to show that all integrals in our formulae can be restricted to a well-defined finite integration region.
A generalized ray is a sharply-peaked Gaussian, thus in practice we ignore its tails and assume that $\psi_{\beta,\rho}(\va{r}) = 0$ when $\abs*{\va{r}-\va{r}_0}>\varrho$, for some $\varrho>0$.
Clearly, this can be done to arbitrary precision: the tail mass of $\psi_{\beta,\rho}$ decays rapidly as a function of $\varrho$.
Therefore, all integrations are over the support of a generalized ray: in the sourcing (\cref{measurement_at_detector_element_backward,MC_sourcing}), measurement (\cref{measurement_operator}), and interaction operator acting upon a generalized ray (\cref{interaction_operator}), viz. $\mathpzc{K}^{\timereversal}\{g\}$, formulae are confined to the (finite) spatial extent of $\psi_{\beta,\rho}$.

We will now show that we may always restrict the integration that defines an arbitrary interaction kernel $K^{\timereversal}$ to a finite region.
Consider the action of that kernel (\cref{interaction_operator}) on a generalized ray $g$:
\begin{align}
    \wvd \triangleq& 
        \mathpzc{K}^{\timereversal}\qty{
            g_{\beta,\rho}\qty(\va{r}^\prime,\va{k}^\prime \argsep \va{r}_0, \va{k}_0)
        }
    \nonumber \\
    =&
        \int 
            \dd{\va{r}^\prime} \dd{\va{k}^\prime} 
            K^{\timereversal}\qty(\va{r},\va{r}^\prime,\va{k},\va{k}^\prime) 
            g_{\beta,\rho}\qty(\va{r}^\prime,\va{k}^\prime \argsep \va{r}_0, \va{k}_0)
    \label{locality_integral_optical_response0}
    ~.
\end{align}
Substitute the definition of the kernel (\cref{WDF_kernel_def}) and rewrite the generalized ray $g$ using the definition of the WDF (\cref{def_WDF}) via its wave function $\psi_{\beta,\rho}$ (\cref{def_gen_ray}).
With some basic algebra and proper variable changes the above becomes 
\begin{align}
    \wvd =
        \int 
            \dd{\va{y}^\prime} \dd{\va{y}^\dprime} \dd{\va{x}_o} 
                &\psi_{\beta,\rho}^\star\qty(\va{y}^\prime)
                h\qty(\va{y}^\prime,\va{r}+\tfrac{1}{2}\va{x}_o)
        \nonumber \\[-.75ex]
        \times&
                \psi_{\beta,\rho}\qty(\va{y}^\dprime)
                h^\star\qty(\va{y}^\dprime,\va{r}-\tfrac{1}{2}\va{x}_o)
                \ee^{-\ii \va{k}\cdot\va{x}_o}
    \label{locality_integral_optical_response}
    ~.
\end{align}
That is, the system's optical response function $h(\va{r}_o,\va{r}_i)$ only contributes to the result $\wvd$ when $\psi_{\beta,\rho}(\va{r}_o)$ is non-negligible.
Apply the relation $h(\va{r}_o,\va{r}_i)=h^\star(\va{r}_i,\va{r}_o)$ to \cref{locality_integral_optical_response}, and we also deduce that $h(\va{r}_o,\va{r}_i)$ only contributes when $\psi_{\beta,\rho}(\va{r}_i)$ is non-negligible.
Hence, we may always replace the optical response function $h$ with 
\begin{align}
    \bar{h}\qty(\va{r}_o,\va{r}_i)
    \triangleq&
        \begin{cases}
            h\qty(\va{r}_o,\va{r}_i) &\qif* \abs{\va{r}_o}<\varrho \text{ and } \abs{\va{r}_i}<\varrho \\
            0 &\qotherwise*
        \end{cases}
\end{align}
in the definition of a diffraction kernel (\cref{WDF_kernel_def}), thereby limiting the integration to the spatial extent $\varrho$ of the incident generalized ray, while remaining accurate to arbitrarily good precision.
We have shown that our formalism achieves weak-locality.

Compare the above with related work:
If we were to replace the generalized rays with plane waves, viz. $\psi_{\beta,\rho}\propto\mathrm{exp}(-\ii\va{k}\cdot\va{r})$, locality would no longer be recoverable as the integration in \cref{locality_integral_optical_response} must happen over the entire domain.
Similarly, consider a Wigner-based perfectly-local formalism (for example \citet{Cuypers_Haber_Bekaert_Oh_Raskar_2012}) where we set $g=\dirac(\va{r}-\va{r}_0)\dirac(\smash{\va{k}-\va{k}_0})$.
Then, \cref{locality_integral_optical_response0} becomes $\wvd=K^{\timereversal}$, and the integration region in the definition of $K$ (\cref{WDF_kernel_def}) can no longer be constrained (without sacrificing linearity).

\paragraph{Completeness}

Given a photoelectric detector, whose detectable states (its WDF) can be written as \cref{wdf_coherent_states}, our derivations in \cref{section_backward_light_transport} are exact.
We stress that essentially all detectors of interest work via the process of photoelectric detection \cite[Chapters 3 and 4]{Leonhardt1997-xz}.
Arbitrary detector geometry and detection properties, quantified by $\mathcal{D}$, are supported.
Interactions of the WDF with the scene via a diffraction kernel $K$, as in \cref{wdf_convolution_kernel,WDF_kernel_def}, is a general formalism \cite{testorf2010phase}, and no restrictions are placed on the total interaction operator $\mathpzc{K}^{\timereversal}$.
Likewise, the sourcing distribution $\wvd[s]$ can be arbitrary.

Because (multivariate) Gaussians serve as an overcomplete functional basis, an arbitrary Husimi Q distribution can be written as a finite superposition of Gaussians to arbitrarily good precision.
We discuss this further in our supplemental material.
Therefore, the recursive light transport process, formalised by \cref{rendering_equation}, is well defined.
The restriction of a generalized ray, as well as the definition of the interaction kernel (\cref{WDF_kernel_def}), to a finite spatial region can also be done to an arbitrarily good precision.
Therefore, our formalism is complete: able to reproduce any wave-optical effect observable by a photoelectric detector.

\subsection{Linearity of Generalized Rays} \label{subsection_linearity_of_generalized_rays}

Consider Young's iconic double-slit interferometer (illustrated in \cref{fig_twin_slit_a}), and assume we use a coherent laser source. 
This experiment will be used to 
\begin{enuminline}
    \item lend insight into how generalized rays are always able to maintain linearity, even when the incident illumination is perfectly coherent; and,
    \item numerically validate our formalism (see \cref{section_numeric_validation}).
\end{enuminline}
In addition, while our practical interactive rendering algorithm that we introduce later (\cref{section_interactive_rendering}) neglects free-space diffractions, this experiment demonstrates that our formalism may reproduce such effects.

As the coherent light passes through the slits, it diffracts around the slits, resulting in a (coherent) superposition phasor $\varphi$ at the screen.
However, we do not observe $\varphi$ at singular points, but only over regions, ``pixels'', on the screen (blue line on the screen in \cref{fig_twin_slit_a}).
The spatial extent of these pixels must be positive, due to the uncertainty relation \cite{mandel1995optical}: we are never able to resolve light at infinite resolution.
Because interference is averaged out over the spatial extent of a pixel, oscillations of $\varphi$ that are more rapid than the pixel's extent do not contribute to observable interference effects.

Let points $\va{s}_{1,2}$ located on the slits act as point sources (i.e., the Huygens–Fresnel principle).
Given identical peak amplitudes, their superposition phasor at a point $\va{r}$ on the screen can be written as 
\begin{align}
    \varphi
        &\approx
        \tfrac{
            \ee^{-\ii k z}
        }{z}
        \qty[
            \exp(-\ii \tfrac{k}{z} \va{s}_1^\superperp\cdot\va{r}^\superperp)
            +
            \exp(-\ii \tfrac{k}{z} \va{s}_2^\superperp\cdot\va{r}^\superperp)
        ]
    ~,
\end{align}
where $k$ is the wavenumber, as before, the superscript $\perp$ denotes projection upon the plate plane ($z=0$), and we made the typical Fraunhofer (far-field) approximation \cite{Born1999principles} (note, we only make this approximation for the illustrative analysis here, the results in \cref{fig_twin_slit_b,fig_twin_slit_plots} were obtained using exact formulae).
The observed intensity of this phasor is then proportional to
\begin{align}
    \abs{\varphi}^2
        &\propto
        \tfrac{2}{z^2}\qty(
            1
            +
            \cos\qty[\tfrac{k}{z} \qty(\va{s}_1^\superperp-\va{s}_2^\superperp)\cdot\va{r}^\superperp]
        )
    \
    \label{two_point_interference}
    ~.
\end{align}
The interference term above oscillates with an angular frequency of $k\tfrac{l_s}{z}$ (as a function of screen position $\va{r}^\superperp$), where $l_s=\abs*{\va{s}_1^\superperp-\va{s}_2^\superperp}$ is the distance between the points on the slits.

\input{fig_twin_slit.tex}

\input{fig_twin_slit_plots.tex}

\input{fig_twin_slit_plots_cuypers.tex}

We now make contact with sampling theory: By the Shannon sampling theorem, we may resolve that interference term without aliasing, only if we observe the phasor over a spatial extent (i.e., a pixel) no greater than $\tfrac{1}{2k(l_s/z)}$.
Under the setting of our small-angle approximation, $\flatfrac{l_s}{z}$ approximates the angle between $\va{s}_1,\va{s}_2$ subtended from $\va{r}$, denoted $\theta_s$.
Denote the spatial extent of a pixel as $l_r$,and we establish the \emph{sampling relation}: $l_r (k\theta_s) \leq \half$.
This relation resembles the uncertainty relation \cite[Chapter 4]{mandel1995optical} but with the inequality reversed ($l_r$ is the spatial variance and $k\theta_s$ is the angular variance scaled by the wavenumber).

When the angular extent $\theta_s$ between the interference sources $\va{s}_{1,2}$ is large enough to violate the sampling relation above, i.e. $l_r (k\theta_s) > \half$, the observer is no longer able to resolve the interference term.
Generalized rays at the detector ($\rho=0$) are minimum-uncertainty constructs, i.e. fulfil the equality in the sampling relation.
Therefore, generalized rays quantify exactly the spatial and angular extent over which interference effects may be resolved.
The arguments above are not limited to double-slit diffraction, and apply, in general, to any superposition from two or more point sources.
To conclude: it is the \emph{integration over the observer's spatial extent that induces decoherence}, allowing generalized rays to regain linearity under any illumination conditions.

We illustrate that decoherence in \cref{fig_twin_slit_b}, where we perform the experiment with increasing slit distances.
The standard deviation of the spatial extent of a generalized ray (illustrated with dashed cyan lines in \cref{fig_twin_slit_a}) at the slits is about $100\lambda$, i.e. a full width at half maximum of $\sim\kern-1ex 235\lambda$.
While the characteristic double-slit interference pattern is visible at first (when $d$ is smaller than the spatial extent of a generalized ray), it slowly dissolves into a pair of (mutually-incoherent) single-slit diffraction patterns. 
Observe the increasing frequency of the secondary fringes in \cref{fig_twin_slit_b} with increasing $d$.

\subsection{Numeric Validation} \label{section_numeric_validation}

We perform the described double-slit experiment using a pair of methods:
\begin{enuminline}
    \item forward transport, where we diffract the incident plane wave directly via the exact Rayleigh-Sommerfeld (RS) explicit diffraction integral of the first kind \cite[Chapter 3.2]{mandel1995optical}, and integrate over the detector (each pixel); and 
    \item backward transport using our formalism, i.e. sampling with generalized rays and diffracting them through the slits over their spatial extent.
\end{enuminline}
The absolute difference between the two methods is shown in the inset at the bottom right of each pattern in \cref{fig_twin_slit_b} (intensity is normalized such that the peak fringe of each pattern has intensity of 1).
The differences (due to ignoring generalized rays' tail) are negligible.

Once the distance $d$ becomes larger than the spatial extent of a generalized ray that reaches the slits, generalized rays no longer solve a double-slit diffraction problem.
Indeed, they don't need to: interference between the two slits does arise at singular points, but is not observable over a pixel's extent.
This can be seen in \cref{fig_twin_slit_plots}, where for large $d$ we can see that the rapidly-oscillating double-slit interference pattern arises everywhere, but is integrated out over the spatial extent of a pixel.
In contrast to generalized rays, explicit diffraction integrals (first method above) have no means of quantifying the extent over which observable interference may arise.
This means that the explicit method needs to (wastefully) integrate each sample over each pixel and over the entire plate.
Therefore, our method (second method) is an order-of-magnitude faster than the explicit method (\SI{8.9}{\second} compared to \SI{246}{\second}). 
This experiment demonstrates that generalized rays achieve well-defined weak locality.

Other methods suffer from similar deficiencies: they fail to derive machinery that is able to quantify the extent over which weak-locality can be maintained.
For example, \citet{Cuypers_Haber_Bekaert_Oh_Raskar_2012} describe a WDF-based formalism with perfectly local primitives.
However, as discussed in \cref{section_related_work}, such a formalism forgoes either locality or linearity: formulating a light-matter interaction requires integration over the entire scene, as defined by the interaction kernel (\cref{WDF_kernel_def}).
In \cref{fig_twin_slit_cuypers_plots} we show that if we were to limit the spatial extent of integration of an interaction kernel, while insisting on linearity (ignore the bilinear superposition term in \cref{wdf_bilinear}), incorrect results are produced.
Indeed, no wave-optical formalism may be simultaneously perfectly local and linear.


\section{Interaction Kernels} \label{section_interaction_kernels_examples}

Developing a light-matter interaction operator entails deriving an analytic method to compute the diffracted distribution $\husimiQ[o]$ which arises from the interaction operator acting on an arbitrary generalized ray, and express it as a finite sum of generalized rays, viz. \cref{MC_light_transport_Q1}.
Formally, this can always be done with arbitrarily high accuracy (as discussed in \cref{section_theory}).
However, doing so in practice can be analytically-involved, and at times application-specific assumptions are made in order to simplify the analysis. 
Often, an importance sampling strategy, that is used to sample a generalized ray out of the sum in \cref{MC_light_transport_Q1}, is desired.
In this Section we present a few important example interactions.

Because time-reversal is equivalent to wavevector reversal and phase conjugation \cite[Chapter 2.3]{Geru_2018}, we may rewrite the action of an arbitrary time-reversed interaction operator on a generalized ray (that enters the rendering equation \cref{rendering_equation}) as 
\begin{align}
    \mathpzc{K}^{\timereversal}\qty{
        g_{\beta,\rho}\qty(\va{r}^\prime,\va{k}^\prime \argsep \va{r}_0, \va{k}_0)
    }
    =&
        \mathpzc{K}\qty{
            g_{\beta,\rho}\qty(\va{r}^\prime,\va{k}^\prime \argsep \va{r}_0, -\va{k}_0)
        }
    \label{action_time_reversal}
\end{align}
(recall that the WDF is real).
Instead of deriving time-reversed interactions, we only need to reverse the generalized ray's wavevector.

We classify light-matter interactions into two types:
\begin{enumerate}
    \item \textsc{Simple} ---
            These arise with \emph{linear optical systems} (also known as ``ABCD systems''): propagation through a homogeneous medium with a slowly-varying refractive index (including free space), and reflection or refraction at a smooth interface.
            Under simple interactions the dynamics are identical to ray-optical dynamics (see our supplemental material).
    \item \textsc{Diffractive} ---
            All other interactions, e.g., scattering by a rough surface:
            there are many ways to formalise a diffractive interaction, and we will discuss a few examples.
\end{enumerate}
Generalized rays are unique in simultaneously
\begin{enuminline}
    \item behaving like classical rays under simple interactions; \item and being \emph{Wigner-representable} \cite{torre2005linear}, i.e. they admit well-defined wave functions.
\end{enuminline}
The first point only holds for constructs that are fully defined by their \nth{2}-order moments matrix (like a generalized ray, which is a Gaussian beam), while the second point only applies to phase-space constructs that fulfil the uncertainty relation.
For example, the second point does not hold for perfectly-local Wigner-based formalisms (the WDF $\wvd\equiv\dirac(\va{r})\dirac(\va{k})$ admits no wave function).

\paragraph{Free-space propagation (simple interaction)}

Let a generalized ray be parameterized by its mean spatial position $\va{r}_0$, mean wavevector $\smash{\va{k}}_0$, and $\beta,\rho$, as before.
Using its phase-space representation (\cref{symplectic_Gaussian}), observe that we may express its spatial and wavevector variances as $\sigma_r^2=\beta^2/2$, and $\sigma_k^2=(1+\rho^2)/(2\beta^2)$, respectively.
$\sigma_r$ quantifies the spatial extent occupied by the generalized ray, and $\sigma_k/k_0$ the solid angle into which the generalized ray propagates.

Let $z>0$ be the distance of propagation, and $\bar{z}=z/\abs*{\smash{\va{k}_0}}$ the distance normalized by the mean wavevector.
Then, the corresponding kernel (\cref{WDF_kernel_def}) is
\begin{align}
    K_{\raisemath{-.125em}{\substack{
        \text{\tiny\scshape f\kern-.25pt r\kern-.25pt e\kern-.25pt e} \\[-.45ex]
        \text{\tiny\scshape s\kern-.25pt p\kern-.25pt a\kern-.25pt c\kern-.25pt e}
    }}}\qty(\va{r},\va{r}^\prime,\va{k},\va{k}^\prime) 
    \triangleq&
        \dirac\qty(\va{k}^\prime - \va{k})
        \dirac\qty(\va{r}^\prime+\bar{z}\va{k} - \va{r})
    ~,
\end{align}
i.e. propagation in direction $\smash{\va{k}}$.
Applying \cref{interaction_operator}, the diffracted (propagated) distribution after interaction becomes
\begin{sizeddisplay}{\small}
\begin{align}
    \!\!
    \husimiQ[o]\qty(\va{r},\va{k})
    =&
        \mathpzc{K}_{\hspace*{.7pt}\raisemath{-.2em}{\substack{
            \text{\tiny\scshape f\kern-.25pt r\kern-.25pt e\kern-.25pt e} \\[-.45ex]
            \text{\tiny\scshape s\kern-.25pt p\kern-.25pt a\kern-.25pt c\kern-.25pt e}
        }}}
        \qty{
            g_{\beta,\rho}\qty(\va{r}^\prime,\va{k}^\prime \argsep \va{r}_0, \va{k}_0)
        }
    =
        g_{\beta_o,\rho_o}\qty(\va{r},\va{k} \argsep \va{r}_0+\bar{z}\va{k}_0, \va{k}_0)
    \label{K_free_space_propagation}
    \\[-.2ex]
    \text{with}\quad&\quad
    \beta_o^2 = \beta^2 + \bar{z} \qty(2\rho + 2\bar{z}\tfrac{1+\rho^2}{2\beta^2}) 
    \qand
    \rho_o^2 = \qty(\rho + 2\bar{z}\tfrac{1+\rho^2}{2\beta^2})^2
    \nonumber
    ~.
\end{align}
\end{sizeddisplay}
The reader may verify that after propagation $\smash{\va{k}_0,\sigma_k^2}$ remain invariant, i.e. the direction and solid angle into which the generalized ray propagates do not change; 
and, the spatial variance transforms as $\sigma_r^2 \to \sigma_r^2 + \order*{\bar{z}^2\sigma_k^2}$, i.e. the space occupied by the generalized ray increases proportionally to the propagation distance times the solid angle.
Finally, the mean position $\va{r}_0$ is shifted by $z$.
That is, all the generalized ray's parameters transform under classical ray-optical dynamics, as would be expected with free-space propagation.

\paragraph{Reflection/refraction at an interface (simple interaction)}

Assume a smooth, flat interface. 
Let $\mathpzc{K}$ be the interaction operator, quantifying the action of reflection or refraction at that interface.
Let $\smash{\va{k}}$ be the mean wavevector of the incident generalized ray, and we denote {\small$\smash{\va{k}_0^\text{(r)}}$} as that wavevector after the reflection (by the law of reflection) or refraction (by Snell's law) at the interface.
Let $a$ be the Fresnel coefficient for the interaction.
Then,
\begin{sizeddisplay}{\small}
\begin{align}
    \husimiQ[o]\qty(\va{r},\va{k})
    =&
        \mathpzc{K}_{\hspace*{.7pt}\raisemath{-.2em}{\substack{
            \text{\tiny\scshape r\kern-.25pt e\kern-.25pt f\kern-.25pt l\kern-.25pt e\kern-.25pt c\kern-.25pt t} \\[-.45ex]
            \text{\tiny\scshape r\kern-.25pt e\kern-.25pt f\kern-.25pt r\kern-.25pt a\kern-.25pt c\kern-.25pt t}
        }}}
        \qty{
            g_{\beta,\rho}\qty(\va{r}^\prime,\va{k}^\prime \argsep \va{r}_0, \va{k}_0)
        }
    =
        a g_{\beta,\rho}\qty(\va{r},\va{k} \argsep \va{r}_0, \va{k}_0^\text{(r)})
    \label{K_reflection_refraction}
    ~.
\end{align}
\end{sizeddisplay}
The parameters $\va{r}_0,\beta,\rho$ are unchanged as no propagation takes place.
That is, with generalized rays this interaction (and other simple interactions) mimics the classical dynamics.
Reflection and refraction are polarization-dependent phenomena. 
Vectorization can be done trivially (generalized ray per each polarization component).
We handle polarization differently, see \cref{section_interactive_rendering}.

\paragraph{Diffraction grating (diffractive interaction)}

A benefit of generalized rays is that they admit a well-defined wave function, $\psi_{\beta,\rho}$.
Hence, to formulate a diffractive interaction one may work in phase space with WDFs, or in position space with wave functions (via a diffraction integral, electromagnetic theory, etc.). 
Furthermore, the generalized ray's wave function is a simple (coherent) Gaussian beam, which has been extensively studied in optical literature.

Apply the Fraunhofer diffraction integral \cite{Born1999principles} to a one-dimensional sinusoidal grating of period $\Lambda$ and height $h$, yielding
\begin{sizeddisplay}{\small}
\begin{align}
    \!\!\!\!
    \husimiQ[o]\qty(\va{r},\va{k})
    =&
        \mathpzc{K}_{\raisemath{-.2em}{\text{\tiny\scshape \kern.5pt g\kern-.25pt r\kern-.25pt t\kern-.25pt n}}}
        \qty{
            g_{\beta,\rho}\qty(\va{r}^\prime,\va{k}^\prime \argsep \va{r}_0, \va{k}_0)
        }
    \approx\!\!
        \sum\nolimits_{j} a_j g_{\beta,\rho}\qty(\va{r},\va{k} \argsep \va{r}_0, \va{k}_0^{(j)})
    \label{K_grating}
    ~,
\end{align}
\end{sizeddisplay}
where $a_j = \besselJ{j}(h k / 2)^2$ is the intensity of the $j$-order diffracted lobe, $\besselJ{j}$ is the Bessel function of \nth{1} kind and 
{\small${\va{k}{}_0^{(j)}}$} is the diffracted wavevector in direction $\sin\theta_o=\sin\theta_i-j\tfrac{\lambda}{\Lambda}$, where $\theta_{i,o}$ are the incident and diffraction directions w.r.t. the grating direction and $\lambda$ is the wavelength.
The diffraction grating also very slightly enlarges the correlation constant $\rho$, but we ignore that effect for simplicity.

\paragraph{Scatter by moderately-rough surface (diffractive interaction)}

We make a simplifying assumption: ignore the beam curvature of a generalized ray, viz. $\psi_{\beta,\rho} \equiv 
        \ee^{\ii \smash{\va{k}_0} \cdot (\va{r}-\va{r}_0)}
        \ee^{-\smash{\tfrac{1}{2\beta^2}} \abs*{\va{r}-\va{r}_0}^2}
$.
This serves to transform the Gaussian beam into a spatially-modulated plane wave, easing the analysis.
We then may apply the Harvey-Shack surface scatter theory \cite{Krywonos2006}: 
\begin{sizeddisplay}{\small}
\begin{align}
    \husimiQ[o]\qty(\va{r},\va{k})
    =&
        \mathpzc{K}_{\raisemath{-.2em}{\text{\tiny\scshape \kern.5pt s\kern-.25pt u\kern-.25pt r\kern-.25pt f\kern-.25pt a\kern-.25pt c\kern-.25pt e}}}
        \qty{
            g_{\beta,\rho}\qty(\va{r}^\prime,\va{k}^\prime \argsep \va{r}_0, \va{k}_0)
        }
    \nonumber \\
    \approx&
        \int\! \dd{\va{k}_0^\text{(r)}}
        f_\text{HS}\qty(\va{k}_0 \to \va{k}_0^\text{(r)}) 
        \ 
        g_{\beta,\rho}\qty(\va{r},\va{k} \argsep \va{r}_0, \va{k}_0^\text{(r)})
    \label{K_surface_HS}
    ~,
\end{align}
\end{sizeddisplay}
where $f_\text{HS}$ is the Harvey-Shack BRDF, quantifying the scattering for incident and diffracted wavevectors $\smash{\va{k}_0}$ and $\smash{\va{k}}_0^\text{(r)}$, respectively.
In practice, sampling generalized rays from the distribution $\husimiQ[o]$ is done via Monte-Carlo integrating the expression above, thereby rewriting it in the form of \cref{MC_light_transport_Q1}:
\begin{align}
    \husimiQ[o]\qty(\va{r},\va{k})
    \approx&
        \frac{1}{J}
        \sum_{j}^{J}
            \tfrac{
                f_\text{HS}\qty(\va{k}_0 \to \bar{\vb{k}}_{0,j}^\text{(r)}) 
            }{
                p_j
            }
            g_{\beta,\rho}\qty(\va{r},\va{k} \argsep \va{r}_0, \bar{\vb{k}}_{0,j}^\text{(r)})
\end{align}
(up to a normalization constant),
where $\bar{\vb{k}}_{0,j}^\text{(r)}$ are the set of $J$ sampled mean scattering directions, and $p_j$ are the sampling probabilities.
In our implementation, the Harvey-Shack BRDF $f_\text{HS}$ is importance sampled using the technique described by \citet{Holzschuch2017}.

In \cref{K_grating,K_surface_HS} we make optical approximations that are reasonable for our target application: rendering at optical frequencies.
One consequence of using the (approximative) Harvey-Shack model for surfaces is that we only model the averaged scatter, where every generalized ray interacts with the entire distribution of surface frequencies.
Therefore, surface imperfections, such as glints, or optical speckle, do not arise.
We stress that the assumptions we made here are not a limitation of our light transport formalism. 
Also note that, in general, $\mathpzc{K}$ may describe cross-wavelength scattering, e.g., due to fluorescence or phosphorescence, however we ignore such effects in our implementation.

Other materials that we use in our rendered scenes are also derived under the assumption that we ignore the curvature of a generalized ray.
Their interaction formula then echoes \cref{K_surface_HS}, but with the BRDF $f_\text{HS}$ replaced with the BRDF of the relevant interaction that acts on plane waves.


\section{Wave-Optical Rendering} \label{section_rendering}

In this Section we introduce our rendering algorithm. 
We start with the accurate algorithm that applies our light transport formalism, as derived in \cref{section_backward_light_transport}, exactly.
Later, in \cref{section_interactive_rendering} we show that with a few reasonable assumptions, our formalism gives rise to an interactive wave-optical rendering algorithm that is easy to implement in a modern path tracer.

Our wave-optical light transport formalism works by propagating backwards, from the detector, generalized rays and simulating their interactions with the scene under time-reversed dynamics, and finally measuring their contributions by integrating over the sourcing distributions of light sources.
A generalized ray is a time-reversed Gaussian beam sourced from the detector:
\cref{def_gen_ray,symplectic_Gaussian} define its wave function and WDF, respectively.

Crucially, generalized rays are always \textbf{linear} (generalized rays never interfere), and \textbf{weakly local}:
we limit the spatial extent of the generalized ray to some finite radius around its mean position $\va{r}_0$.
Being Gaussian beams, with rapidly-decaying tails, this can be easily done to arbitrarily high accuracy.
For example, a radius of $\varrho=2.5\beta$, where $\beta^2/2$ is the generalized ray's spatial variance, captures $>99.8\%$ of its weight (such $\varrho$ is used in \cref{fig_twin_slit}).
Finally, this formalism is \textbf{complete} (further discussed in \cref{section_gr_analysis}).

\input{fig_algorithm.tex}

In \cref{alg_rendering} we summarise our general-purpose, accurate light transport algorithm:
\begin{itemize}
    \item   \textcolor{blue}{\textbf{Sourcing.}}
            A generalized ray is fully described by its mean position $\va{r}_0$, mean wavevector $\va{k}_0$ (quantifying direction of propagation and the wavelength), spatial variance $\beta^2/2$ and correlation parameter $\rho$.
            For each sample, these parameters are sourced via \cref{MC_sourcing}, thereby sourcing a generalized ray initiating a path.
    \item   \textcolor{red!80!black}{\textbf{Propagation.}}
            We propagate the generalized ray through free space, until it (or a part of it) encounters matter.
            As we limit the spatial extent of a generalized ray to a region (ball of radius $\varrho$) around its mean $\va{r}_0$, tracing it reduces to beam (or cone) tracing through the scene.
            We formalize the transformation of the generalized ray's parameters, {\small $\qty{\va{r}_0,\!\smash{\va{k}_0},\!\beta,\!\rho}$}, under free-space propagation in \cref{K_free_space_propagation}.
    \item   \textcolor{green!65!black}{\textbf{Measurement.}}
            If the generalized ray (or part of it) encounters a light source, we integrate the generalized ray over the sourcing distribution, as formalised by the measurement operator (\cref{measurement_operator}).
            When no light source is present, $\mathfrak{I}\equiv0$.
    \item   \textcolor{violet!80!black}{\textbf{Light-matter interaction.}}
            We compute the interaction operator $\mathpzc{K}$ (\cref{interaction_operator}), that quantifies the action of matter that falls within the spatial extent of the beam (ball of radius $\varrho$ centred at $\va{r}_0$).
            This operator acts upon the traced generalized ray (with wavevector reversed, as in \cref{action_time_reversal}), giving rise to the scattered distribution $\husimiQ[o]$ (as in \cref{light_transport_Q1}).
            Then, a diffracted generalized ray is sampled out of this distribution and the light transport simulation continues.
\end{itemize}
The tuple $\vb{\pi}$ is a bookkeeping variable that keeps track of the indices of the generalized rays that were sampled at each step.

The algorithm presented above is a novel wave-optical rendering algorithm.
As formulated, it is able to simulate rigorous wave-optical light transport accurately (to arbitrarily good precision) using weakly-local, linear generalized rays.
This algorithm is employed for the simulation in \cref{fig_twin_slit}.
A pair of practical difficulties may arise:
\begin{enumerate}
    \item   In contrast to classical light transport, propagation cannot be done via ray tracing and requires significantly-slower beam tracing, as generalized rays are not perfectly local (indeed, no wave-optical formalism may be simultaneously perfectly local and linear). 
    \item   Because generalized rays occupy a positive extent, the operator $\mathpzc{K}$ in the \textcolor{violet!80!black}{light-matter interaction} stage may need to account for multiple different materials simultaneously (for example, when a generalized ray is incident upon an edge or a slit, or upon a surface with spatially-varying scattering characteristics).
            This can be understood as the interaction being a composition of two or more operators, viz. $\mathpzc{K}\equiv\mathpzc{K}_1 \composition \mathpzc{K}_2$.
            Formally, this can always be done, but, in practice, formalising the composed interaction and sampling generalized rays from it (thereby accounting for the mutual interference of the scattered generalized rays from all materials that are involved in the interaction) can be difficult.
\end{enumerate}

\input{fig_sample_solve.tex}

An efficient implementation of beam tracing, designed for modern GPUs, as well as deriving some composite interactions of interest (e.g., free-space diffraction of a generalized ray around geometry) would be of great interest, and are left for future work.
Such work would be especially important for wave-optical simulation at longer wavelengths (e.g., RADAR). 
In the rest of this paper our focus is on rendering at optical frequencies.
We present next our interactive wave-optical rendering implementation, where we make a simplifying application-specific assumption.


\input{fig_FC.tex}

\subsection{Sample-Solve} \label{section_sample_solve}

In this Subsection we draw a formal connection between our backward (sensor-to-source) formalism to \emph{optical coherence}.
The purpose of this is to connect our formalism to other forward-based models and computational optics tools, thereby enabling the application of bi-directional techniques.
For our application of interest, we will show that we may use a forward PLT pass as a cheap variance reduction technique.

\paragraph{Relation to optical coherence}

The central quantity in the study of optical coherence \cite{wolf2007introduction} is the \emph{cross-spectral density} (CSD) of a statistical ensemble of light waves, viz.
\begin{align}
    \csd\qty(\va{r}_1, \va{r}_2)
    \triangleq&
        \ev{
            \psi\qty(\va{r}_1)
            \psi^\star\qty(\va{r}_2)
        }
    \label{def_csd}
    ~,
\end{align}
where $\ev{\cdot}$ denotes ensemble-averaging over the wave ensemble, and the wave function $\psi$ is now understood as a realization from that ensemble.
From the definitions of the WDF and the CSD, \cref{WDF_kernel_def,def_csd}, it is easy to observe that the CSD is the Fourier transform of the ensemble-averaged WDF, therefore we may write
\begin{align}
    \csd\qty(\va{r}-\tfrac{1}{2}\va{d},\va{r}+\tfrac{1}{2}\va{d})
    \propto&\ 
        \ev{
            \frft{\wvd(\va{r},\va{k}^\prime)}\qty(\va{d})
        }
    \label{csd_to_wdf}
\end{align}
(up to an irrelevant constant), where the Fourier transform $\frft$ is w.r.t. the primed variable, $\smash{\va{k}^\prime}$.
We may immediately conclude that optical coherence at a fixed point $\va{r}$ only depends on the (ensemble-averaged) light's wavevector distribution in phase space.

Under partial coherence, the observed values are often ensemble-averaged values \cite{wolf2007introduction} (and can be understood as time averaging over the period of detection), i.e. $\ev{I}$.
Then, partial coherence is trivially accounted for by replacing the sourcing WDF $\wvd[s]$ with its ensemble-averaged counterpart, $\ev{\wvd[s]}$, in all our formulae.
$\ev{\wvd[s]}$ is the Fourier transform of the sourced CSD of light (\cref{csd_to_wdf}).

We term the \emph{diffusivity} $\matb{\Omega}$ of a WDF to be the angular variance of propagation from the mean direction of propagation.
The diffusivity $\matb{\Omega}$ is a $2\times 2$ positive-definite matrix $\matb{\Omega}$, allowing for anisotropy.
For example, in the isotropic case, the diffusivity can be written as $\sigma_k^2/k_0^2$, where $\sigma_k^2$ is the distribution's wavevector variance.
The variance in the solid angle into which the bundle propagates is then $\varOmega=\abs*{\matb{\Omega}}$.
Then, from \cref{csd_to_wdf} we may formally derive the following result (see \cref{subsection_rendering_coherence} in our supplemental material):
\begin{align}
    \matb{\Theta} = \lambda^2 \matb{\Omega}^{-1}
    \label{coherence_diffusivity}
    ~,
\end{align}
where $\matb{\Theta}$ is the \emph{shape matrix} from PLT theory \cite{Steinberg_practical_plt_2022}, i.e. the (inverse) \nth{2}-order moments of spatial variance.
Deriving equality relations between higher-order moments is also possible, however we do not require higher-order moments.

A consequence of the above relation between optical coherence and the diffusivity of light is the well-known connection between the coherence area of light, i.e. $\abs*{\matb{\Theta}}$, and the solid angle subtended by a thermal source \cite{mandel1995optical}: $\abs*{\matb{\Theta}}=\tfrac{\lambda^2}{\varOmega}$.
Though note that the relation \cref{coherence_diffusivity} is more general, and establishes a direct connection between optical coherence and light's distribution in the wave-optical phase space, with no assumptions on the light sourcing process or its state-of-coherence at other regions in space.

\paragraph{Sample-solve}

We present a simple two pass algorithm:
first, we \emph{sample} the wave-optical distribution of light using generalized rays, using the rendering algorithm we presented so far, \cref{alg_rendering}. 
The process continues until a generalized ray encounters a light source, thereby a path $\vb{\pi}$ that connects the detector to a light source is found.
The effective diffusivity over that path is well defined.
In-place of the \textcolor{green!65!black}{measurement} stage (line 9 in \cref{alg_rendering}),
we now apply a forward pass: we use PLT machinery \cite{Steinberg_practical_plt_2022} to \emph{solve} for the partially-coherent light transport over the path $\vb{\pi}$.
More formally, instead of integrating over the ensemble-averaged sourcing WDF, $\ev{\wvd[s]}$, we use PLT to source a partially-coherent beam that corresponds to $\ev{\wvd[s]}$ from that light source, and retrace the steps forward (source-to-sensor) taken by the generalized ray over the path.
As the PLT beam captures more information (it is ``wider'' than a generalized ray), this forward solve pass serves as a \emph{variance-reduction technique}, by computing the partially-coherent optical response for the sampled path.
See \cref{fig_FC}.

In our domain of interest---wave-optical rendering---applying PLT makes sense:
The optical coherence of light is a primary factor in limiting our ability to resolve wave-interference effects.
Other applications might find a different \emph{solve} pass to be more appropriate:
For example, integrating optical speckle statistics can be done via a sample-solve approach, where first we \emph{sample} paths connecting the detector to a light source, and then integrate statistics in a \emph{solve} pass.
Such applications are beyond the scope of this paper, however they serve to highlight the generality of this simple sample-solve approach: it bridges a gap between classical path tracing tools and wave optics, via the generalized ray construct, and enables the application of powerful sampling techniques in a wider context.


\subsection{Interactive Rendering} \label{section_interactive_rendering}

In similar manner to optical rendering with PLT \cite{Steinberg_practical_plt_2022}, we make the assumption that the spatial extent $\varrho$ of a generalized ray is smaller or comparable to the scene's smallest geometric details (we refer to the explicit geometry against which we trace rays, and not analytically defined details like surface properties).
Therefore, we may neglect the generalized ray's spatial extent and use ray tracing to propagate it.
Then, the \textcolor{red!80!black}{propagation} stage in our rendering algorithm, \cref{alg_rendering}, reduces to simple ray tracing, and in the \textcolor{violet!80!black}{light-matter interaction} stage we do not consider composite interactions with multiple materials.

The simplifying assumption above is reasonable for our target application: high-performance rendering at optical frequencies.
In terms of optical effects, this assumption means that we neglect free-space diffractions (light ``bending'' around geometry) and interference across distinct materials.
We stress that this assumption is not a limitation of our wave-optical light transport formalism, as presented in \cref{section_theory}.

Under the assumption above, surprisingly few changes are required to convert a classical unidirectional, backward path tracer into a path tracer compatible with wave optics: 
generalized rays replace the classical ray as the ``point queries'' of light's behaviour.
In practice, doing so requires essentially only a few changes:
\begin{enuminline}
    \item transitioning to a spectral, vectorized (polarization-aware) path tracer; 
    \item transforming generalized rays on free-space propagation, as in \cref{K_free_space_propagation}; and
    \item reformulating all the BSDFs in terms of generalized rays, as in \cref{K_surface_HS}.
\end{enuminline}

\input{fig_MS.tex}

\input{fig_bike.tex}

\paragraph{Implementation details}

To \textcolor{blue}{source} a generalized ray from the detector distribution $\mathcal{D}$ of a pixel: the variance $\beta^2/2$ is taken to be a fraction of the pixel spatial size; $\rho=0$ at sourcing; for simplicity, we set the detection efficiency $\alpha=1$; the initial mean position $\va{r}_0$ is sampled uniformly at random on the pixel; and, $-\va{k}_0$ is the (mean) wavevector.
A wavelength $\lambda_0$ (such that $\abs*{\va{k}_0}=\tfrac{2\mpi}{\lambda_0}$) is drawn randomly at uniform from the visible spectrum.
This wavelength drives the path sampling process, hence we refer to it as the ``Hero wavelength'' \cite{Wilkie2014-yg}.

The \textcolor{red!80!black}{propagation} of a generalized ray remains unchanged, as described in \cref{section_rendering} and formalized by \cref{K_free_space_propagation}.
\textcolor{violet!80!black}{Light-matter interactions} are done as described in \cref{section_interaction_kernels_examples}, and we continue to ignore the generalized ray's curvature for diffractive interactions.
Diffractive interaction formulae then take the form of \cref{K_surface_HS}, and we are able to importance sample these interactions (i.e., importance sampling the function $f$ in \cref{K_surface_HS}).

We continue to transform the generalized ray's parameters, viz. {\small $\qty{\va{r}_0,\!\smash{\va{k}_0},\!\beta,\!\rho}$}, on propagation and on interaction with matter.
Therefore, we are able to quantify exactly the spatial extent over which a generalized ray interacts with matter.
Materials that describe deterministic scattering features, like a surface's explicit microgeometry, or a multilayered stack (as seen in the beetle's wings in \cref{fig_teaser}), can be accurately rendered.
This is not possible with any existing linear wave-optical formalisms: none of them are able to recover weak locality, and would require integration over the entire material.

In addition, when a generalized ray encounters matter, we attempt to connect the path to a light source via next-event-estimation (NEE).
We also perform ``Russian roulette'' early path termination.
We note that NEE, ``Russian roulette'', and importance sampling are all standard sampling techniques in path tracing \cite{pharr2016physically}.
We are also able to apply other advanced sampling techniques in our wave-optical rendering algorithm:
To demonstrate that we have implemented manifold sampling \cite{Hanika2015-gv,Zeltner2020-xl}, see \cref{fig_MS}.

As discussed in \cref{section_sample_solve}, the \textcolor{green!65!black}{measurement} stage in \cref{alg_rendering} is replaced with a PLT solve pass.
Every time a path has been connected to a light source, either organically or via NEE, the solve stage kicks in to compute the partially-coherent transport over the sampled path.
If the sampled path does not contain dispersive delta segments (e.g., refraction through a smooth dielectric interface), in addition to the Hero wavelength, we also importance sample 3 additional wavelengths from the light's emission spectrum.
We therefore transport one to four spectral samples per sampled path.
The solve stage applies PLT in an essentially unchanged manner, see \citet{Steinberg_practical_plt_2022} for more details.

Our implementation supports two types of light sources:
\begin{enumerate}
    \item \emph{Distant light sources} ---
        These include analytic distant sources, with a predefined solid angle that they subtend from the scene, as well as environment maps, in which case light is sourced from a small cluster of a few pixels, defining the solid angle.
    \item \emph{Emissive geometry light sources} ---
        where light is sourced from a small area on the emitting triangle.
\end{enumerate}
As a design choice, our implementation ignores polarization during the sample pass (i.e., the backward transport with generalized rays).
This is done for simplicity and performance: decomposing and inverting Mueller matrices, and other polarimetric interactions, can be cumbersome and expensive.
No errors are introduced: the solve pass is fully vectorized.


\subsection{Results} \label{section_results}

Our results are comprised of three main scenes: 
\begin{enumerate}
    \item 
        \textbf{Snake enclosure} (\cref{fig_teaser,fig_snake}).
        This scene is illuminated by multiple light sources: the sun, the sky (diffused sunlight), as well as a pair of industrial \SI{4100}{\kelvin} fluorescent lamps with a decent colour rendering index of 82 located at the back of the enclosure. 
        Sunlight and skylight arrive from the opening at the top.
        This is a difficult scene to render: most of the light that arrives at the different diffractive materials is indirect.
    \item 
        \textbf{Manifold sampling} (\cref{fig_MS}).
        A highly-detailed scene that we use as our manifold-sampling playground.
    \item 
        \textbf{Bike} (\cref{fig_bike}).
        Adapted from \citet{Steinberg_practical_plt_2022} (appearance is not expected to match, as our materials are different).
\end{enumerate}
In addition to the above, the CD scene is used for analysis of partially-coherent sampling, \cref{fig_PC}, and comparison to the state-of-the-art, \cref{fig_mitsuba}.
\cref{fig_snake} demonstrates that wave-optics is a global process: the appearance of a material, and the observable diffractive phenomena, depend on the wave properties of light.
Accurate reproduction of such effects cannot be done at the material level, but instead requires a wave-optical simulation throughout the entire scene, as we do.
Our supplemental material contains additional renderings, as well as animated videos of all these scenes, showcasing the performance of our method.

\input{fig_performance.tex}

\input{fig_PC.tex}

Performance metrics, i.e. rendering resolution and samples-per-pixel (spp) count are given in each figure, and summarised in \cref{table_performance}.
For the paper figures we used high-quality, converged (i.e., very high samples-per-pixel) renders.
Nevertheless, our method and implementation enable interactive wave-optical rendering at 1 spp, and the frame times for interactive rendering are also given in the figures and in \cref{table_performance}.
Similar to other modern GPU-accelerated path tracers, we use a denoiser for interactive rendering.
This enables generating acceptable images at 1 spp, and allows the user to interact with and edit the scene in real-time.
The videos in our supplemental material were also rendered using a denoiser.

Low-spp images of the bike scene, with and without the denoiser, are shown in \cref{fig_bike} (bottom).
Low sample count is sufficient for the vast majority of the materials, including the diffractive birefringent dielectrics.
An exception is the dispersive diffraction lobes, which are visible on the floor, and arise due to the diffraction-grated wheel brake surface.
This is due to a couple of reasons: 
\begin{enuminline}
    \item 
        A diffraction grating scatters into many wavelength-dependent lobes, hence requires a moderate amount of spectral samples.
    \item 
        These are rendered via manifold sampling (MS), as discussed in \cref{fig_MS}.
        However, MS is only initiated during the sample stage when a path is scattered from the floor into the grating.
        The probability of finding such a connection organically is rather low (roughly about 1 in a few dozen), as the grated surface is quite small.
\end{enuminline}
The difficulty of rendering the diffracted lobes of a diffraction grating is also analysed in \cref{fig_MS} (bottom).
This problem is similar to the classical problem of rendering dispersive caustics, e.g., on the bottom of a pool with a disturbed water surface; a problem where classical unidirectional path tracers struggle as well.

\paragraph{Comparison with the state-of-the-art}

We compare our interactive wave-optical renderer to PLT \cite{Steinberg_practical_plt_2022}.
PLT employs a bi-directional path tracer, and samples partially-coherent BSDFs.
As discussed, because the coherence of light cannot be known a priori when tracing paths backwards (i.e., the sampling problem), they set a global lower limit on the coherence of light.
Partially-coherent BSDFs are then sampled with respect to that limit.
We emulate such partially-coherent sampling in our renderer, and compare the performance in the simple CD scene, see \cref{fig_PC}.
The difference is dramatic: we observe over a thousand-fold increase in convergence performance when sampling with generalized rays compared to partially-coherent sampling.

\input{fig_mitsuba.tex}

We also compare by rendering the same CD scene directly with PLT, see \cref{fig_mitsuba}.
An increase in performance is expected, as our implementation is GPU accelerated, nevertheless, we record up to a several thousand speed gain for a similar-quality rendering (for image areas that admit diffractive materials).
Our equal-sample convergence performance is increased by a factor of 1 to 8.
This is despite the fact that their path tracer is bi-directional (ours is unidirectional) and each sample propagates 64 uniformly sampled spectral samples (compared to the up to 4 per sample with our renderer), therefore they do a considerably greater amount of work per sample.
\cref{fig_PC,fig_mitsuba} highlight the sampling problem and emphasize the need for better solutions: the state-of-the-art is able to achieve decent results when diffractive materials are directly illuminated, making good use of a bi-directional path-tracer.
However, it struggles greatly when it isn't trivial to connect paths from a light source to these materials.


\section{Conclusion} \label{section_conclusion}

The primary contribution in this work is our backward (sensor-to-source) accurate wave-optical light transport formalism.
The derivations in \cref{section_theory} are exact.
Our formalism is \emph{complete}, subject to photoelectric detectors: meaning any wave-optical phenomena that may be observed by a photoelectric detector can be simulated to arbitrarily good precision, with light of optical and non-optical frequencies, and of any state of coherence or polarization.
At the core of our formalism is the generalized ray: \emph{weakly-local} and \emph{linear} constructs that sample the distribution of light from the detector.
In \cref{fig_twin_slit} we numerically validate this formalism, as well as delineate the highly-general process by which generalized rays are able to regain linearity in all conditions: only interfering phasors that can be resolved over the spatial extent of the detection element, without aliasing, contribute to observable interference effects.

Some practical considerations are left for future work: 
\begin{enuminline}
    \item efficient beam tracing; and, 
    \item sampling the interaction of a generalized ray with a composite interaction operator, like free-space diffraction through a slit.
\end{enuminline}
These considerations are especially important for rendering and light transport with longer, non-optical wavelengths, where the spatial extent of a generalized ray is significantly larger than at optical frequencies.

As a sample application, in this paper we focus on wave-optical rendering in the visible spectrum.
A reasonable, application-specific simplification is made: like PLT, we assume that the spatial extent of the generalized ray is small compared to the scene's geometric details.
By doing so, we have made it possible for the first time to efficiently render complex scenes under a highly-accurate wave optical model. 
Our novel sample-solve approach makes it possible to apply decades of work in path tracing sampling algorithms to wave optics rendering, now allowing the construction of paths from the camera and the use of advanced sampling techniques like manifold next-event estimation in the wave-optical setting. 
Our algorithms are highly efficient and are amenable to GPU implementation; we show interactive one sample per pixel rendering of complex scenes, on a modern GPU, at a performance that is orders-of-magnitude faster than the state-of-the-art.
Our implementation is available with our supplemental material.

\input{fig_snake.tex}

We stress: no wave-optical formalism may be perfectly local and linear simultaneously.
We then sacrifice perfect locality in favour of weak locality, while retaining linearity.
Even in our practical, interactive rendering algorithm, where we make the simplifying assumption discussed above, we are able to accurately quantify the spatial extent over which a generalized ray interacts with matter.
This is crucial for an efficient light transport simulation: without weak locality, light-matter interactions need to be integrated over the entire scene.
As discussed, any forward (source-to-sensor) weakly-local, linear formalism inevitably suffers from the \emph{sampling problem}, where weak locality cannot be established when tracing backward (sensor-to-source).
Our formalism is unique in providing such machinery that is always able to regain weak locality and linearity under backward light transport.

A generalized ray can be understood as the wave-optical analogue of the classical ray:
Generalized rays traverse free space, trace Eikonals through media with a slowly-varying refractive index, and reflect or refract at a smooth interface between media in an identical fashion to classical rays.
A generalized ray is the most compact construct permissible under wave optics.


\bibliographystyle{ACM-Reference-Format}
\bibliography{paper}


\appendix


\end{document}

%% file: preamble.tex
\usepackage{graphicx}
\usepackage[export]{adjustbox}
\usepackage{grffile}
\usepackage{booktabs}
\usepackage[dvipsnames]{xcolor}
\usepackage{tcolorbox}

\usepackage{cellspace}
\usepackage{makecell}
\usepackage{tabularx}
\usepackage{ragged2e}
\usepackage{multicol}

\usepackage{subcaption}
\usepackage{afterpage}
\usepackage{dblfloatfix}

\usepackage{etoolbox}
\usepackage{xparse}
\usepackage{xifthen}
\usepackage{scalerel}
\usepackage{stackengine}

\usepackage{fancyhdr}

\usepackage{upgreek}
\usepackage{bbm}
\usepackage{thmtools}
\usepackage{mathtools}
\usepackage{amsmath}
\usepackage{amsfonts}

\usepackage{xfrac}
\usepackage{interval}
\usepackage{physics}
\usepackage{tensor}
\usepackage{derivative}

\usepackage[linesnumbered,ruled,vlined]{algorithm2e}

\usepackage{relsize}
\usepackage{scalefnt}
\usepackage[inline]{enumitem}
\usepackage[outline]{contour}
\usepackage[normalem]{ulem}

\usepackage[super]{nth}

\usepackage{tikz}
\usepackage{tikz-3dplot}
\usepackage{tkz-euclide}
\usepackage{pgfplots}

\usepackage{xr-hyper}
\usepackage{hyperref}
\usepackage[nameinlink,capitalize]{cleveref}
\crefname{subsection}{Subsection}{Subsections}
\Crefname{subsection}{Subsection}{Subsections}
\crefname{Table}{Table}{Tables}
\Crefname{Table}{Table}{Tables}

\usepackage{siunitx}

\usepackage{nomencl}

\tcbuselibrary{skins}
\tcbsetforeverylayer{shield externalize}

\usetikzlibrary{angles,arrows,arrows.meta,calc,quotes,intersections,shapes,positioning,scopes,shadings,fadings}
\usetikzlibrary{decorations.markings,decorations.text,decorations.pathmorphing,decorations.pathreplacing,patterns,patterns.meta}
\usetikzlibrary{3d,spy}
\usetikzlibrary{pgfplots.colormaps}
\usetikzlibrary{external}
\pgfplotsset{compat=newest}
\usepgfplotslibrary{fillbetween}
\usepgfplotslibrary{colormaps}
\usepgfplotslibrary{groupplots}
\usepgfplotslibrary{units}
\pgfplotsset{
	colormap/magma/.style={%
		/pgfplots/colormap={magma}{%
			rgb=(0.001462, 0.000466, 0.013866)
			rgb=(0.035520, 0.028397, 0.125209)
			rgb=(0.102815, 0.063010, 0.257854)
			rgb=(0.191460, 0.064818, 0.396152)
			rgb=(0.291366, 0.064553, 0.475462)
			rgb=(0.384299, 0.097855, 0.501002)
			rgb=(0.475780, 0.134577, 0.507921)
			rgb=(0.569172, 0.167454, 0.504105)
			rgb=(0.664915, 0.198075, 0.488836)
			rgb=(0.761077, 0.231214, 0.460162)
			rgb=(0.852126, 0.276106, 0.418573)
			rgb=(0.925937, 0.346844, 0.374959)
			rgb=(0.969680, 0.446936, 0.360311)
			rgb=(0.989363, 0.557873, 0.391671)
			rgb=(0.996580, 0.668256, 0.456192)
			rgb=(0.996727, 0.776795, 0.541039)
			rgb=(0.992440, 0.884330, 0.640099)
			rgb=(0.987053, 0.991438, 0.749504)
		},
	},
	colormap/inferno/.style={%
		/pgfplots/colormap={inferno}{%
			rgb=(0.001462, 0.000466, 0.013866)
			rgb=(0.037668, 0.025921, 0.132232)
			rgb=(0.116656, 0.047574, 0.272321)
			rgb=(0.217949, 0.036615, 0.383522)
			rgb=(0.316282, 0.053490, 0.425116)
			rgb=(0.410113, 0.087896, 0.433098)
			rgb=(0.503493, 0.121575, 0.423356)
			rgb=(0.596940, 0.154848, 0.398125)
			rgb=(0.688653, 0.192239, 0.357603)
			rgb=(0.775059, 0.239667, 0.303526)
			rgb=(0.851384, 0.302260, 0.239636)
			rgb=(0.912966, 0.381636, 0.169755)
			rgb=(0.956852, 0.475356, 0.094695)
			rgb=(0.981895, 0.579392, 0.026250)
			rgb=(0.987464, 0.690366, 0.079990)
			rgb=(0.973088, 0.805409, 0.216877)
			rgb=(0.947594, 0.917399, 0.410665)
			rgb=(0.988362, 0.998364, 0.644924)
		},
	},
	colormap/plasma/.style={%
		/pgfplots/colormap={plasma}{%
			rgb=(0.050383, 0.029803, 0.527975)
			rgb=(0.186213, 0.018803, 0.587228)
			rgb=(0.287076, 0.010855, 0.627295)
			rgb=(0.381047, 0.001814, 0.653068)
			rgb=(0.471457, 0.005678, 0.659897)
			rgb=(0.557243, 0.047331, 0.643443)
			rgb=(0.636008, 0.112092, 0.605205)
			rgb=(0.706178, 0.178437, 0.553657)
			rgb=(0.768090, 0.244817, 0.498465)
			rgb=(0.823132, 0.311261, 0.444806)
			rgb=(0.872303, 0.378774, 0.393355)
			rgb=(0.915471, 0.448807, 0.342890)
			rgb=(0.951344, 0.522850, 0.292275)
			rgb=(0.977856, 0.602051, 0.241387)
			rgb=(0.992541, 0.687030, 0.192170)
			rgb=(0.992505, 0.777967, 0.152855)
			rgb=(0.974443, 0.874622, 0.144061)
			rgb=(0.940015, 0.975158, 0.131326)
		},
	},
	colormap/inferno,
	colormap/plasma,
	colormap/magma,
}

\DeclareSIUnit\erg{erg}

\intervalconfig{ soft open fences }

\DeclareMathAlphabet\mathcalcmsy{OMS}{cmsy}{m}{n}
\DeclareMathAlphabet\mathbfcalcmsy{OMS}{cmsy}{b}{n}
\DeclareMathAlphabet\mathbfcal{OMS}{cmsy}{b}{n}
\DeclareMathAlphabet\mathbfscr{OMS}{mdugm}{b}{n}
\DeclareMathAlphabet\mathantt{OMS}{antt}{r}{n}
\DeclareMathAlphabet\mathbfantt{OMS}{antt}{b}{n}
\DeclareMathAlphabet\mathzapf{T1}{pzc}{mb}{it}
\DeclareMathAlphabet\mathpzc{OT1}{pzc}{m}{it}

\newcommand{\raisemath}[1]{\mathpalette{\@raisemath{#1}}}
\newcommand{\@raisemath}[3]{\raisebox{#1}{$#2#3$}}
\newcommand{\scalemath}[1]{\mathpalette{\@scalemath{#1}}}
\newcommand{\@scalemath}[3]{\scalebox{#1}{$#2#3$}}

\newcommand\ii{\mathrm{i}}
\newcommand\mpi{\uppi}
\newcommand\ee{\mathrm{e}}

\let\oldva\va
\renewcommand\va[1]  {\oldva{\boldsymbol #1}}

\let\oldvb\vb
\renewcommand\vb[1]  {\oldvb{\boldsymbol #1}}

\let\oldvu\vu
\renewcommand\vu[1]  {\oldvu{\boldsymbol #1}}

\newcommand\mat[1]    {{\boldsymbol #1}}
\newcommand\matb[1]   {\pmb{#1}}

\newcommand\matcalcmsy[1] {\mathbfcalcmsy{#1}}

\newcommand*\superperp{{\mathpalette\@superperp{}}}
\newcommand*\@superperp[2]{\raisebox{-0.3ex}{\scalebox{.8}{$\m@th#1\perp$}}}
\newcommand*\superparallel{{\mathpalette\@superparallel{}}}
\newcommand*\@superparallel[2]{\raisebox{-0.3ex}{\scalebox{.8}{$\m@th#1\parallel$}}}
\newcommand*\subperp{{\mathpalette\@subperp{}}}
\newcommand*\@subperp[2]{\raisebox{0.2ex}{\scalebox{.8}{$\m@th#1\perp$}}}
\newcommand*\subparallel{{\mathpalette\@subparallel{}}}
\newcommand*\@subparallel[2]{\raisebox{0.2ex}{\scalebox{.8}{$\m@th#1\parallel$}}}

\newcommand\overdarrow[2]{\ensurestackMath{%
	\savestack{\tmpbox}{\stretchto{%
		\scaleto{%
			\scalerel*[\widthof{\ensuremath{#1}}]{\kern.9pt\leftrightarrow\kern.4pt}%
					  {\rule[-\textheight/2]{1ex}{\textheight}}
		}{\textheight}%
	}{.66ex}}%
	\stackon[#2]{#1}{\tmpbox}%
}}
\newcommand*\dvec[1]{\overdarrow{#1}{.45pt}}

\newcommand*\transpose{{\mathpalette\@transpose{}}}
\newcommand*\@transpose[2]{\raisebox{\depth}{$\m@th#1\intercal$}}
\DeclareDocumentCommand\conj{g}{\IfNoValueTF{#1}{\star}{#1^\star}}
\newcommand*\composition{\raisebox{0.5ex}{\scalebox{.375}{$\bigcirc$}}}

\DeclareDocumentCommand\rect{g}{\IfNoValueTF{#1}{\operatorname{rect}}{\fbraces{}{}{\operatorname{rect}}{#1}}}
\DeclareDocumentCommand\circ{g}{\IfNoValueTF{#1}{\operatorname{circ}}{\fbraces{}{}{\operatorname{circ}}{#1}}}
\DeclareDocumentCommand\besselJ{m}{\ensuremath{\mathrm{J}_{\mkern-2mu#1}}}

\DeclareMathOperator\dirac{\delta}
\DeclareDocumentCommand\kdelta{g}{\IfNoValueTF{#1}{\operatorname{\dirac}}{\operatorname{\dirac}_{#1}}}

\DeclarePairedDelimiterX{\definp}[2]{\langle}{\rangle}{#1\,\delimsize\vert\,#2}
\def\inp{\@ifstar{\definp}{\definp*}}

\DeclareDocumentCommand\ensemble{sm}{
	\IfBooleanTF{#1}{\langle#2\rangle}{\left\langle#2\right\rangle}
}
\DeclareDocumentCommand\ensemblew{sm}{
	\IfBooleanTF{#1}{\langle#2\rangle_\omega}{\left\langle#2\right\rangle_\omega}
}
\DeclareDocumentCommand\tavg{sm}{
	\IfBooleanTF{#1}{\langle#2\rangle_\textrm{t}}{\left\langle#2\right\rangle_\textrm{t}}
}
\DeclareDocumentCommand\realization{o m}{\IfNoValueTF{#1}{\tensor*{#2}{}}{\tensor*[^{#1}]{#2}{}}}

\DeclareDocumentCommand\mathoperator{m m m}{
	\def\op{\mathchoice%
		{\displaystyle 		\scalebox{1.15}{$#1$}}%
		{\textstyle 		\scalebox{1.0} {$#1$}}%
		{\scriptstyle 		\scalebox{0.95}{$#1$}}%
		{\scriptscriptstyle	\scalebox{0.9} {$#1$}}%
	}
	\def\sub{\mathchoice%
		{\displaystyle 		#3}%
		{\textstyle 		#3}%
		{\scriptstyle 		\scalebox{.9}{$#3$}}%
		{\scriptscriptstyle	\scalebox{.8}{$#3$}}%
	}
	\tensor*{\op}{%
		_{\IfNoValueTF{#3}{}{\mkern-1mu\sub}}
		^{\IfNoValueTF{#2}{}{#2}}
	}
}
\DeclareDocumentCommand\fourieroperator{m m}{\mathoperator{\mathscr{F}}{#1}{#2}}
\DeclareDocumentCommand\frft{s o o g}{%
	\def\op{\fourieroperator{#2}{#3}}
	\def\body{#4 \vphantom{\fourieroperator{}{}}}
	\IfNoValueTF{#4}{\fourieroperator{#2}{#3}}{
		\IfBooleanTF{#1}
			{\op\lbrace\body\rbrace}
			{\fbraces{\lbrace}{\rbrace}{\op}{\body}}
	}%
}

\DeclareDocumentCommand\wvd{o}{
	\mathscr{W}_{\IfNoValueTF{#1}{}{\mkern-1.2mu#1}}
	\IfNoValueTF{#1}{\mkern2mu}{}
}
\DeclareDocumentCommand\husimiQ{o}{
	\mathscr{Q}_{\IfNoValueTF{#1}{}{\mkern-.5mu#1}}
}
\DeclareDocumentCommand\conv{o}{
    \tensor*{\ast}{_{\IfNoValueTF{#1}{}{\mkern-3.5mu#1}}}
}

\DeclareDocumentCommand\laplacianp{m o d()}{ 
	\IfNoValueTF{#1}{%
		\IfNoValueTF{#2}{%
            \IfNoValueTF{#3}{\nabla^{\prime 2}}{\fbraces{\lparen}{\rparen}{\nabla^{\prime 2}}{#3}}%
        }{\fbraces{\lbrack}{\rbrack}{\nabla^{\prime 2}}{#2} \IfNoValueTF{#3}{}{(#3)}}%
    }{\nabla^{\prime 2} #1 \IfNoValueTF{#2}{}{[#2]} \IfNoValueTF{#3}{}{(#3)}}%
}
\DeclareDocumentCommand\gradientp{m o d()}{ 
	\IfNoValueTF{#1}{%
		\IfNoValueTF{#2}{%
            \IfNoValueTF{#3}{\vnabla^\prime}{\fbraces{\lparen}{\rparen}{\vnabla^\prime}{#3}}%
        }{\fbraces{\lbrack}{\rbrack}{\vnabla^\prime}{#2} \IfNoValueTF{#3}{}{(#3)}}%
    }{\vnabla^\prime #1 \IfNoValueTF{#2}{}{[#2]} \IfNoValueTF{#3}{}{(#3)}}%
}

\newcommand\argsep{\ {\scalebox{1.2}{$\boldsymbol ;$}}\ } 

\newmuskip\pFqmuskip
\newcommand*\pFq[6][8]{%
    \begingroup %
    \pFqmuskip=#1mu\relax
    \mathcode`\,=\string"8000
    \begingroup\lccode`\~=`\,
    \lowercase{\endgroup\let~}\pFqcomma
    \tensor*[_#2]{F}{_{#3}}
    \ifthenelse{\isempty{#4}}{}{\qty[\genfrac..{0pt}{}{#4}{#5} \argsep #6]}%
    \endgroup
}
\newcommand*\nFq[5][7]{%
    \begingroup %
    \pFqmuskip=#1mu\relax
    \mathcode`\,=\string"8000
    \begingroup\lccode`\~=`\,
    \lowercase{\endgroup\let~}\pFqcomma
    \tensor*[_#2]{F}{_{#3}}
    \ifthenelse{\isempty{#4}}{}{\qty[#4 \argsep #5]}%
    \endgroup
}
\newcommand\pFqcomma{\mskip\pFqmuskip}

\newcommand*\dprime{{\prime\prime\mkern-1.2mu}}

\newcommand\prettyfrac[2]{\fontfamily{ppl}\selectfont\sfrac{#1}{#2}}
\newcommand\half[0]{\prettyfrac{1}{2}}

\newcommand\quarter[0]{\prettyfrac{1}{4}}
\newcommand\threequarters[0]{\prettyfrac{3}{4}}

\newcommand{\hardsec}{$\scalebox{1.2}{$\boldsymbol{\ast}$}$}

\newenvironment{sizeddisplay}[1]
	{\par\nopagebreak#1\noindent\ignorespaces}
	{\nopagebreak\ignorespacesafterend}

\newenvironment{enuminline}
    {\begin{enumerate*}[label=(\roman*)]}
    {\end{enumerate*}}

\newlist{properties}{enumerate}{1}
\setlist[properties]{label=\textbf{(\Roman*)},
                     ref=(\Roman{*})}
\crefname{propertiesi}{property}{properties}
\Crefname{propertiesi}{Property}{Properties}

\newlist{thmlist}{enumerate}{1}
\setlist[thmlist]{label=(\roman{thmlisti}),
                  ref=\thetheorem.(\roman{thmlisti}),
                  noitemsep}
\Crefname{thmlisti}{Theorem}{Theorems}
\addtotheorempostheadhook[theorem]{\crefalias{thmlisti}{listthm}}

\newlist{deflist}{enumerate}{1}
\setlist[deflist]{label=(\roman{deflisti}),
                  ref=\thetheorem.(\roman{deflisti}),
                  noitemsep}
\Crefname{deflisti}{Definition}{Definitions}
\addtotheorempostheadhook[definition]{\crefalias{deflisti}{listdef}}

\DeclareMathOperator\csd{\mathcalcmsy{C}}
\newcommand\SP{\va{S}}

\DeclareDocumentCommand\SX{g}{
	\SP_{\scalebox{.8}{c}}\IfNoValueF{#1}{\qty(#1)}
}
\DeclareDocumentCommand\gSP{so}{
	\IfBooleanTF{#1}
		{\mathcalcmsy{S}}
		{\vphantom{\widehat{\matcalcmsy{S}}} \smash[t]{\dvec{\matcalcmsy{S}}}}
		^{\refframe{
			\IfNoValueTF{#2}{\vb{\mu}}{#2}
		}}
}
\DeclareDocumentCommand\genrad{so}{
	\IfBooleanTF{#1}
		{\lppscalar}
		{\vphantom{\widehat{\lpp}} \smash[t]{\overdarrow{\lpp}{.3pt}}}
		^{\refframe{
			\IfNoValueTF{#2}{\vb{\mu}}{#2}
		}}
}

\DeclareDocumentCommand\lsmM{sg}{
	\bar{\mat{m}}
	\IfNoValueTF{#2}{}{\IfBooleanTF{#1}{(#2)}{\qty(#2)}}
}

\DeclareDocumentCommand\anisoGaussian{o}{
	\IfNoValueTF{#1} {\mathpzc{g}}
					 {\mathpzc{g}^{\raisemath{-.55ex}{#1}}}
}

\DeclareDocumentCommand\refframe{sg}{
	\IfBooleanTF{#1} {{[\!#2\!]}}
					 {{\raisemath{-.1em}{\scalemath{.825}{[\!#2\!]}}}}
}
\DeclareDocumentCommand\refframesub{sg}{
	\IfBooleanTF{#1} {{[\!#2\!]}}
					 {{\raisemath{.05em}{\scalemath{.825}{[\!#2\!]}}}}
}

%% file: fig_teaser.tex
\begin{teaserfigure}
    \centering
	\tikzset{external/optimize=true}
	\tikzsetnextfilename{teaser\imagesmode}%
	\begin{subfigure}{\teasersize\linewidth}%
        \makebox[\textwidth][c]{\resizebox{1\linewidth}{!}{\begin{tikzpicture}[
            ]
        \end{tikzpicture}}}
	\end{subfigure}%
    \vspace*{\teasercaptionoffset}
    \caption{
        \textbf{From ray optics to wave optics.}
        A scene rendered with our technique, showcasing various wave effects: (a) a Bornite ore with an interfering layer of copper oxide; (b) a Brazilian Rainbow Boa, whose scales are biological diffraction-grated surfaces; and (c) a Chrysomelidae beetle, whose colour arises due to multilayered interference reflectors in its elytron.
        The practical contribution of this paper is the ability to render such complex scenes, under rigorous wave-optical light transport, at a performance that surpasses the state-of-the-art by orders-of-magnitude.
        The objective of this paper is not the appearance reproduction of some material (a ``diffractive BRDF''), but an accurate formulation of wave-optical light transport, where light is rigorously treated as waves globally throughout the entire scene.
        We propose the \emph{generalized ray}: an extension of the classical ray to wave optics.
        The generalized ray retains the defining characteristics of the ray-optical ray: \emph{locality} and \emph{linearity}.
        These properties allow the generalized ray to serve as a ``point query'' of light's behaviour, the same purpose that the classical ray fulfils in rendering.
        Generalized rays enable the application of backward (sensor-to-source) light transport and sophisticated sampling techniques, which are impossible with the state-of-the-art. 
        We indicate resolution and samples-per-pixel (spp) count in all figures rendered using our method.
        While these figures showcase converged (high spp) results, our implementation also allows interactive rendering of all these scenes at 1 spp.
        Frame times (at 1 spp) for interactive rendering are indicated.
        See our supplemental material for the implementation, and additional renderings and videos.
    }
	\Description{}
	\label{fig_teaser}
\end{teaserfigure}

%% file: fig_sampling_problem.tex
\begin{figure}[t]%
    \centering
    \tikzset{external/optimize=true}%
    \tikzsetnextfilename{ray_sampling_problem}%
    \resizebox{.75\linewidth}{!}{\begin{tikzpicture}[
        ]
    \end{tikzpicture}}
    \Description{}%
    \vspace*{-1mm}
    \caption{
        \textbf{The sampling problem.}
        Existing light transport formalisms, like partially-coherent light transport, work by evolving light's properties from the source, through the scene, until light is sensed by a detector.
        Such formalisms are inherently incompatible with backward (sensor-to-source) models of light transport:
        they are not able to formulate a light-matter interaction (a BSDF) without knowledge of light's wave properties, however these properties depend on the light source and evolve throughout the scene, hence are very difficult to predict or estimate in a backward model \cite{Steinberg_practical_plt_2022}.
        Fundamental path tracing and sampling techniques, like importance sampling of interactions, cannot be applied, greatly hampering the practicality and ability of these formalisms to work with complex, real-world scenes.
        Solving this \emph{sampling problem}, i.e. devising a formalism of backward wave-optical light transport, where a wide-range of sampling techniques can be applied, is the primary motivation for this paper.
    }%
    \label{fig_sampling_problem}
    \vspace*{-1mm}
\end{figure}
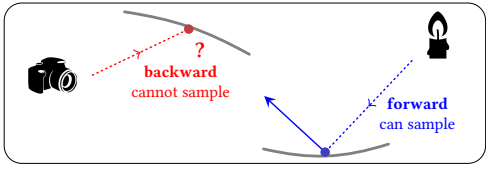

%% file: fig_twin_slit.tex
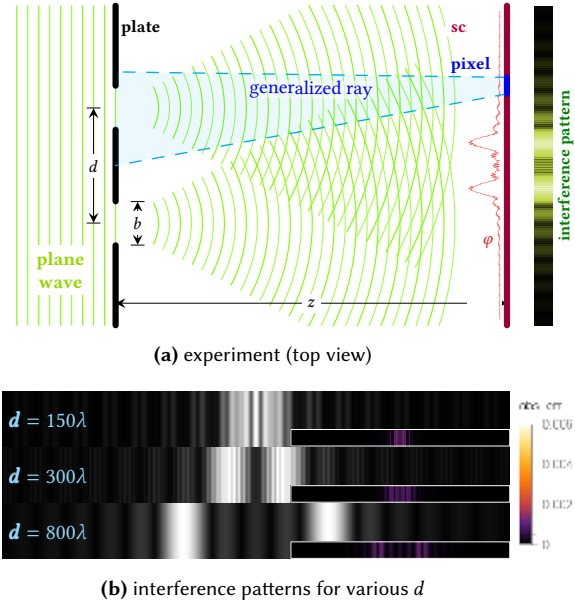
\begin{figure}[t]%
    \centering
    \tikzset{external/optimize=true}%
    \tikzsetnextfilename{twin_slit}%
    \resizebox{.90\linewidth}{!}{\begin{tikzpicture}[
        ]
    \end{tikzpicture}}
    \centering{\phantomsubcaption\label{fig_twin_slit_a}\phantomsubcaption\label{fig_twin_slit_b}}
    \Description{}%
    \caption{
        \textbf{Diffraction through double slits.}
        (a) Schematic of Young's double slit experiment. 
        A pair of slits, of width $b$ and spaced a distance $d$ apart, are cut in a thin, conductive plate.
        A coherent plane wave (illustrated in green) diffracts through the slits, and is observed upon a screen, placed at a distance $z$ from the plate.
        The superposition of coherent light from both slits results in a rapidly-oscillating phasor $\varphi$ (illustrated in red), producing an interference pattern.
        (b) The experiment is performed with increasing slit distances $d$, and we compare our method (sampling incident light with generalized rays) with a ground truth (explicitly diffracting the plane wave through the slits).
        Differences are plotted in the insets at the bottom right of each pattern (intensity of peak fringe is 1). 
        The experiment was performed with wavelength $\lambda=1$ (arbitrary units), $z=10000\lambda$ and $b=40\lambda$. 
    }%
    \label{fig_twin_slit}
    \vspace*{-1mm}
\end{figure}

%% file: fig_twin_slit_plots.tex
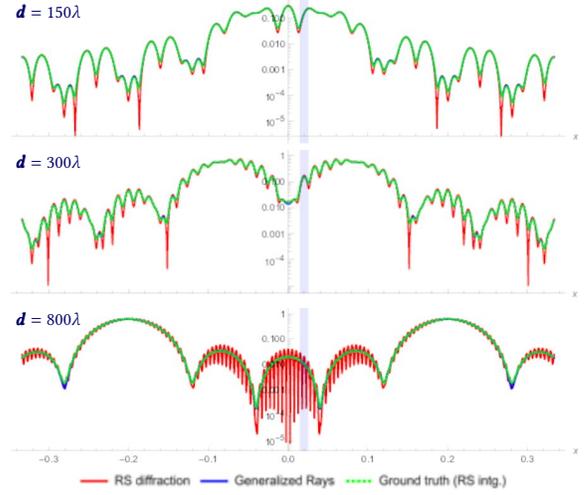
\begin{figure}[t]%
    \centering
    \tikzset{external/optimize=true}%
    \tikzsetnextfilename{twin_slit_plots}%
    \resizebox{.88\linewidth}{!}{\begin{tikzpicture}[
        ]
    \end{tikzpicture}}
    \vspace*{-0.5mm}
    \Description{}%
    \caption{
        Plots of light intensity as a function of $x$ (position on screen) of the experiment in \cref{fig_twin_slit}.
        In red we plot the exact Rayleigh-Sommerfeld (RS) diffraction, i.e. the \emph{unobservable} interference that arises in singular points.
        The spatial extent of a detector on the screen (a pixel in each pattern in \cref{fig_twin_slit_b}) is illustrated as a cyan bar.
        Integrating the RS diffraction over that spatial extent of a pixel computes the numeric ground truth, plotted in dashed green.
        Results obtained with generalized rays are plotted in blue.
    }%
    \label{fig_twin_slit_plots}
    \vspace*{-1.5mm}
\end{figure}

%% file: fig_twin_slit_plots_cuypers.tex
\begin{figure}[t]%
    \centering
    \tikzset{external/optimize=true}%
    \tikzsetnextfilename{twin_slit_plots_cuypers}%
    \resizebox{.88\linewidth}{!}{\begin{tikzpicture}[
        ]
    \end{tikzpicture}}
    \Description{}%
    \caption{
        \textbf{Loss of locality with \citet{Cuypers_Haber_Bekaert_Oh_Raskar_2012}.}
        Red plot (exact RS diffraction) is as in \cref{fig_twin_slit_plots}.
        Let $\rho_\text{wbsdf}$ be the radius of integration of a diffraction kernel (\cref{WDF_kernel_def}) in \citet{Cuypers_Haber_Bekaert_Oh_Raskar_2012}.
        Limiting that integration radius produces erroneous results, hence \emph{locality is entirely lost} with their method: correctness requires integration over the entire scene ($\rho_\text{wbsdf}=\infty$).
    }%
    \label{fig_twin_slit_cuypers_plots}
    \vspace*{-1mm}
\end{figure}
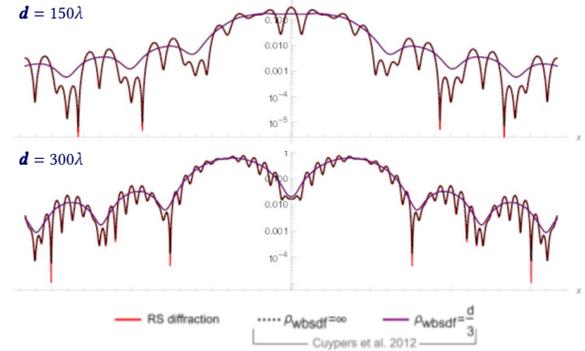

%% file: fig_algorithm.tex
{
\SetKwComment{Comment}{/* }{ */}
\SetCommentSty{footnotesize}
\SetAlFnt{\small}
\SetAlgoNlRelativeSize{-2}

\begin{algorithm}[t]
    \caption{
        \textbf{Our rendering algorithm.}
    }
    \label{alg_rendering}

    \ForEach{detector (pixel) $\mathcal{D}$}{
        $I \gets 0$%
        \Comment*[r]{will hold the computed detected intensity}

        \For(\Comment*[f]{sample the backward light transport}){$n \gets 1$ to $N$}{ 
            {\color{blue}{\small $\qty{\va{r}_0,\!\va{k}_0,\!\beta,\rho}$}$\gets$ source generalized ray}%
            \Comment*[r]{\cref{MC_sourcing}}


            $\alpha \gets \tfrac{\alpha^{(n)}}{N p^{(n)}} $%
            \Comment*[r]{path weight; p is sampling probability}

            \While{path depth $<$ max depth}{
                { \color{red!80!black}propagate the generalized ray {\small $\qty{\va{r}_0,\!\va{k}_0,\!\beta,\!\rho}$}}\;

                \BlankLine
                \Comment{measurement, or solve pass (\cref{section_sample_solve})}
                { \color{green!65!black}\small 
                $I \gets I + \alpha\cdot\mathfrak{I}\qty{
                    g_{\beta,\rho}\qty(\va{r}^\prime,\va{k}^\prime \argsep \va{r}_0, \va{k}_0)
                }$}%
                \Comment*[r]{\cref{measurement_operator}}

                \BlankLine
                \Comment{light-matter interaction}

                {\color{violet!80!black}
                $\mathpzc{K} \gets $ interaction operator around $\va{r}_0$%
                }%
                \;
                {\color{violet!80!black}\small
                $\husimiQ[o] \gets \mathpzc{K}\qty{
                    g_{\beta,\rho}\qty(\va{r}^\prime,\va{k}^\prime \argsep \va{r}_0, -\va{k}_0)
                }$%
                }%
                \Comment*[r]{rendering equation}

                {\color{violet!80!black}
                {\small $\qty{\va{r}_0,\!\va{k}_0,\!\beta,\!\rho}$} $\gets$  sample $\husimiQ[o]$ 
                }%
                \Comment*[r]{\cref{MC_light_transport_Q1}}

                \BlankLine

                $\alpha \gets \alpha \tfrac{\alpha^{\vb{\pi} + (n_m)}}{N_m}$%
                \Comment*[r]{update weight ($\alpha^{\vb{\pi} + (n_m)}, N_m$ as in \cref{MC_light_transport_Q1})}

            }
        }
    }
\end{algorithm}
}

%% file: fig_sample_solve.tex
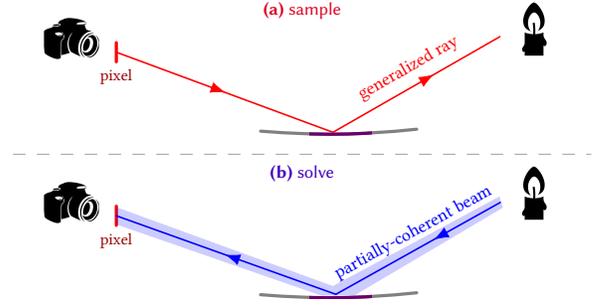
\begin{figure}[t]%
    \centering
    \tikzset{external/optimize=true}%
    \tikzsetnextfilename{sample_solve}%
    \resizebox{.9\linewidth}{!}{\begin{tikzpicture}[
        ]
    \end{tikzpicture}}
    \vspace*{-2mm}
    \Description{}%
    \caption{
        \textbf{Sample-solve.}
        Our path tracing algorithm (a) uses generalized rays (dotted lines) to \emph{sample} paths through the scene.
        Generalized rays are always linear, therefore classical sampling techniques apply essentially unchanged.
        Once a path is sampled (solid red path), we (b) \emph{solve} for the partially-coherent light transport, by applying PLT \cite{Steinberg_practical_plt_2022} from the light source to the sensor across all intermediate interactions.
    }%
    \label{fig_sample_solve}
    \vspace*{-1mm}
\end{figure}

%% file: fig_FC.tex
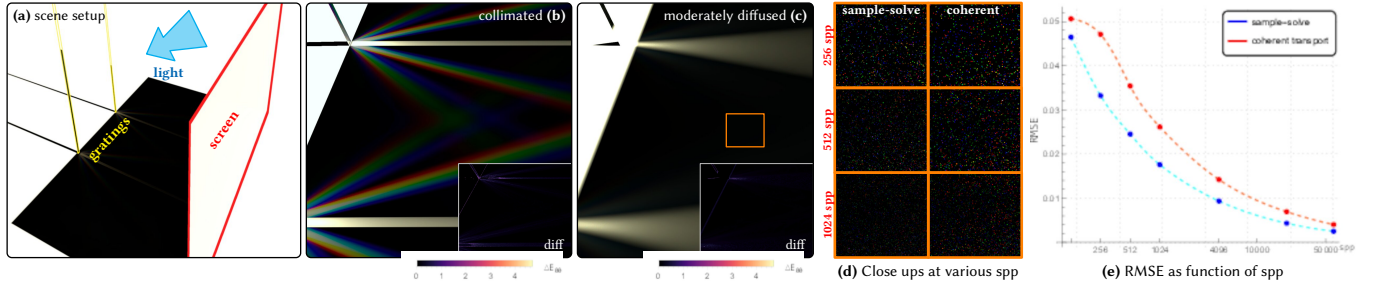
\begin{figure*}[t]%
    \tikzset{external/optimize=true}%
    \tikzsetnextfilename{FC}%
    \resizebox{1\linewidth}{!}{\begin{tikzpicture}[
        ]
    \end{tikzpicture}}
    \vspace*{-4mm}
    \Description{}%
    \caption{
        \textbf{Sample-solve: PLT as a variance reduction technique.}
        During the solve step, we apply PLT to solve for the partially-coherent light transport over a sampled light path.
        (a) Light arrives from the right, illuminating a simple scene. A screen (on the right, red outline) shadows a rectangular area. 
        On the left a pair of thin diffraction gratings (yellow outline) reflect and disperse light.
        (b) When the incident light is highly collimated (subtends a very small solid angle from the scene), hence is moderately coherent, the diffraction lobes are clearly visible.
        (c) As we increase the light's diffusivity (increase its solid angle), the reflection from the direct contribution lobe spreads out, while diffraction lobes mostly disappear, as expected.
        As the only light that arrives to the shadowed region is from the diffraction gratings, this is a good scene to study the benefits of the solve stage, and to do so we render the scene with only fully-coherent transport (no PLT applied).
        (b-c, insets) Difference images between the partially-coherent and fully-coherent sample-solve show that indeed both converge to an identical result.
        (d) However, close ups on the region outlined in orange in (c) show that, while the diffraction lobes are no longer visible, they still induce error, which the partially-coherent solve stage serves to reduce.
        (e) Plot of error in that area as function of sample count suggests that fully-coherent transport requires about 2-5 times the sample count to achieve similar-quality renderings.
    }%
    \label{fig_FC}
\end{figure*}

%% file: fig_MS.tex
\begin{figure}[t]%
    \centering
    \tikzset{external/optimize=true}%
    \tikzsetnextfilename{MS}%
    \resizebox{1\linewidth}{!}{\begin{tikzpicture}[
        ]
    \end{tikzpicture}}
    \Description{}%
    \caption{
        \textbf{Manifold sampling.}
        Example application of an advanced sampling technique.
        When performing next event estimation (NEE), manifold sampling (MS) \cite{Hanika2015-gv,Zeltner2020-xl} enables finding a light path between a surface and a light source across one or more dielectric interfaces.
        In the rendered scene, such a sampled path, that refracts through the dispersive prism (outlined in blue), is visualized via the dotted blue line.
        We also employ MS for NEE on specular reflections: note the reflections off the paint brush's metal handle and off the paint tubes, as well as the thin diffraction grating (outlined in yellow) dispersing light into multiple diffraction lobes. 
        See full high-resolution rendering in our supplemental material.
        (inset) Colour-coded difference compared to a without-MS rendering.
        (bottom) The diffraction grating lobes at increasing samples-per-pixel (spp).
    }%
    \label{fig_MS}
\end{figure}
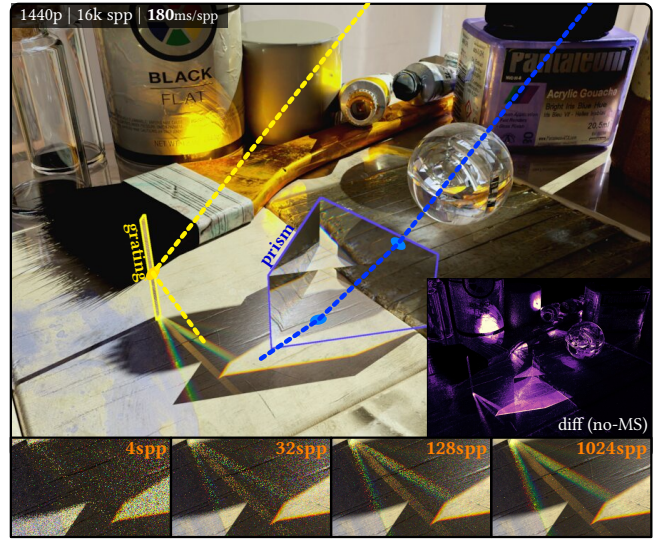

%% file: fig_bike.tex
\begin{figure*}[t]%
    \centering
    \tikzset{external/optimize=true}%
    \tikzsetnextfilename{bike}%
    \resizebox{.99\linewidth}{!}{\begin{tikzpicture}[]
    \end{tikzpicture}}
    \vspace*{-2mm}
    \Description{}%
    \caption{
        \textbf{Bike scene.}
        The bike scene from \citet{Steinberg_practical_plt_2022} rendered using generalized rays.
        The scene contains a few materials with visible wave-interference effects, most notably the birefringent dielectrics, (a) like the plastic wheel spoke guard, (b-c) as well as the diffraction grated wheel brake surface.
        (b) when illuminated by direct sunlight, this grating disperses light into visible diffraction lobes; however, (c) when sunlight passes through a diffuser (like clouds), only the direct lobe is visible.
        (bottom) Low spp renderings, with (left side) and without (right side) a denoiser, showcasing the interactive rendering performance of our wave-optical renderer.
    }%
    \label{fig_bike}
\end{figure*}
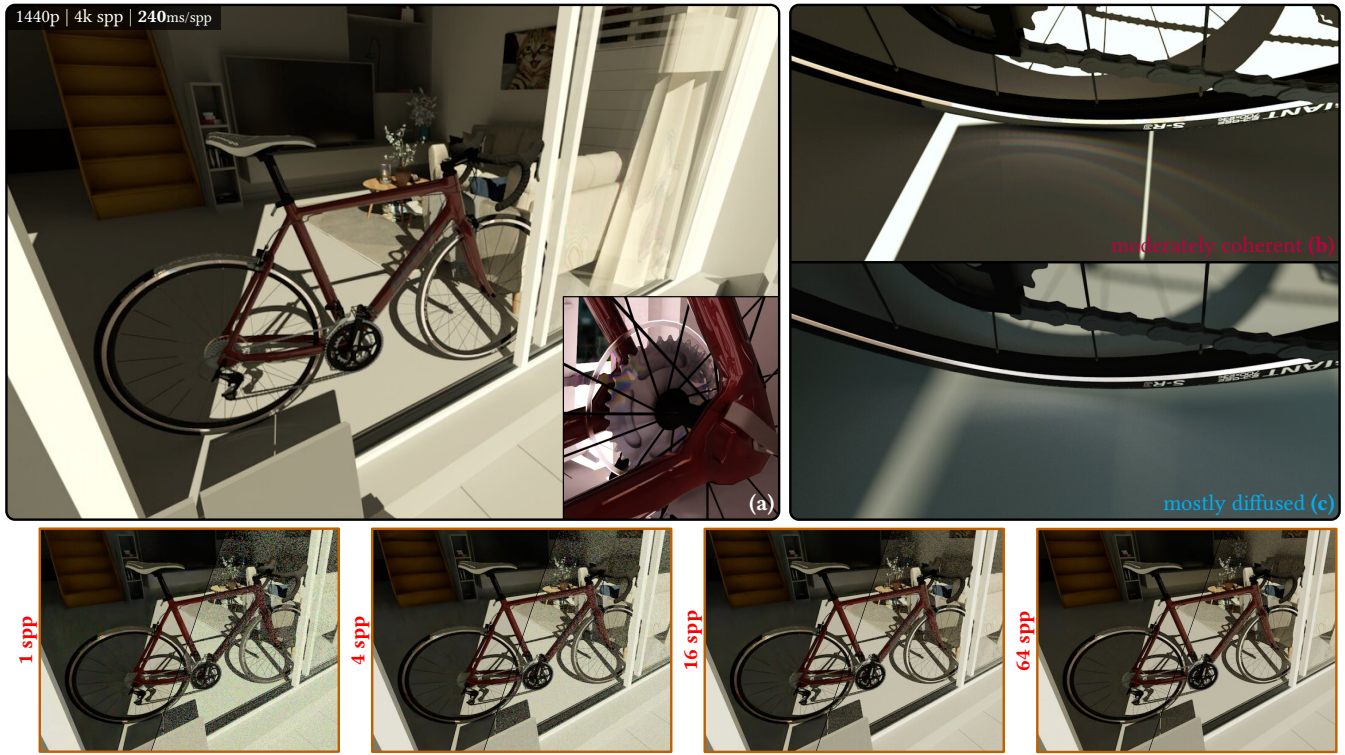

%% file: fig_performance.tex
\begin{table}[b]
	\centering
	\Description{}
    \caption{
			\textbf{Rendering performance}.
			Listed are the resolution and sample count used to generate the figures, as well as the interactive (1 spp) rendering frame times (at the indicated resolution).
			All rendering was done on a NVIDIA\textsuperscript{\textregistered} GeForce RTX\texttrademark\ 3090 GPU.
        }
	
	\newcolumntype{L}{>{\raggedright\arraybackslash}X}%
	\newcolumntype{s}[1]{>{\hsize=#1\hsize}X}%
	\newcolumntype{e}[1]{>{\raggedright\hsize=#1\hsize}X}%
	\begin{tabularx}{0.985\linewidth}%
		{ 
			L@{} c l l 
		}
		\toprule
            \textsf{Scene} & 
            \textsf{Frame time} &
            \textsf{Figure} & 
			{}
			\vspace*{-1.5mm}
			\\
			 & \footnotesize(1 spp) & &
		\\
		\midrule
		    \textbf{\color{violet} Snake enclosure}
			&
			{\color{purple} \SI{116}{\milli\second}}
			&
			\cref{fig_teaser,fig_snake}
			&
			\small 1440p, \emph{64k \textsc{spp}}
		\\
		    \textbf{\color{violet} MS playground}
			&
			{\color{purple} \SI{180}{\milli\second}}
			&
			\cref{fig_MS}
			&
			\small 1440p, \emph{16k \textsc{spp}}
		\\
		    \textbf{\color{violet} Bike}
			&
			{\color{purple} \SI{240}{\milli\second}}
			&
			\cref{fig_bike}
			&
			\small 1440p, \emph{4k \textsc{spp}}
		\\
		    \textbf{\color{violet} CD}
			&
			{\color{purple} \SI{42}{\milli\second}}
			&
			\cref{fig_PC}
			&
			\small 1080p, \emph{2M \textsc{spp}}
		\\
        \bottomrule
	\end{tabularx}
	\label{table_performance}
\end{table}

%% file: fig_PC.tex
\begin{figure*}[t]%
    \centering
    \tikzset{external/optimize=true}%
    \tikzsetnextfilename{PC}%
    \resizebox{1\linewidth}{!}{\begin{tikzpicture}[
        ]
    \end{tikzpicture}}
    \vspace*{-4mm}
    \Description{}%
    \caption{
        \textbf{Partially-coherent sampling.}
        (a) A scene contains a compact disk (CD) that rests next an open CD case, upon which another closed CD case is placed.
        A ceiling-mounted light source illuminates the scene.
        While simple, the scene admits interesting light transport.
        (b) We emulate partially-coherent (PC) sampling of BSDFs, in an identical manner to the state-of-the-art, which manifests the sampling problem: 
        the diffraction lobes are very sharp lobes that are difficult to sample when doing backward path tracing (and the coherence properties of light are unknown) using existing tools.
        The close ups show the area marked in orange rendered at various sample counts.
        Note the sharp difference in noise between PC sampling and our proposed light transport formalism with generalized rays.
        (c) Plot of noise as function of sample count, quantifying the drastic improvement in sampling performance: sampling using generalized rays reduces sample count by a factor of about 4000 for similar-quality rendering.
        (a, inset) A difference image between the two sampling strategies. 
        With the exception of the very high-frequency diffraction effects, the differences are minor, and are due to the PC rendering never converging.
    }%
    \label{fig_PC}
\end{figure*}
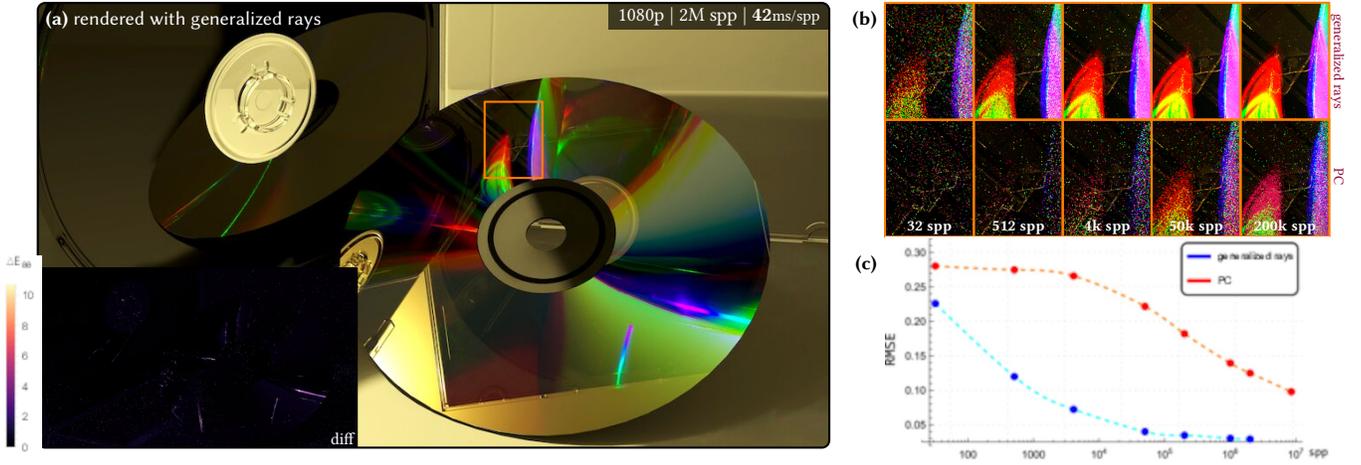

%% file: fig_mitsuba.tex
\begin{figure*}[t]%
    \centering
    \tikzset{external/optimize=true}%
    \tikzsetnextfilename{mitsuba}%
    \resizebox{1\linewidth}{!}{\begin{tikzpicture}[
        ]
    \end{tikzpicture}}
    \Description{}%
    \caption{
        \textbf{Comparison to the state-of-the-art.}
        We render the CD scene using the PLT bidirectional path tracer (BDPT) \cite{Steinberg_practical_plt_2022}.
        Because the illumination of the diffractive CD surface is indirect, convergence is poor, as to be expected given the analysis in \cref{fig_PC}.
        The image was rendered with \num{56000} samples over about \num{315} hours, on an Intel\textsuperscript{\textregistered} Core\texttrademark\ i9-10980XE 18-core CPU.
        (a-c) Their renderer struggles with capturing the high-frequency details of the diffraction grating.
        Furthermore, because they propagate a fixed number (64) of spectral samples, sampled uniformly, clear banding artefacts are reproduced, suggesting that these high-frequency details will fail to converge due to spectral aliasing.
        (right) To analyze and compare the convergence performance to our approach, we plot error per sample and error per rendering-time graphs for the three different regions outlined in red:
        (d) The light transport that arrives to this area (red dotted line) is dominated by light that is not diffracted by the CD.
        Therefore, the per-sample convergence performance of their BDPT renderer (which performs much more work per sample) is significantly superior, as expected.
        (e-f) On the other hand, these regions do sample the CD and the convergence performance of their renderer is considerably inferior.
        A performance improvement is expected, as our renderer is GPU accelerated.
        The performance at region (d), plotted with open circle markers, should then be considered as baseline: up to 100 times faster, depending on sample count.
        The plots pertaining to the diffractive regions (filled markers) show a vastly greater improvement of about \num{1000} to \num{10000} times faster convergence.
    }%
    \label{fig_mitsuba}
\end{figure*}
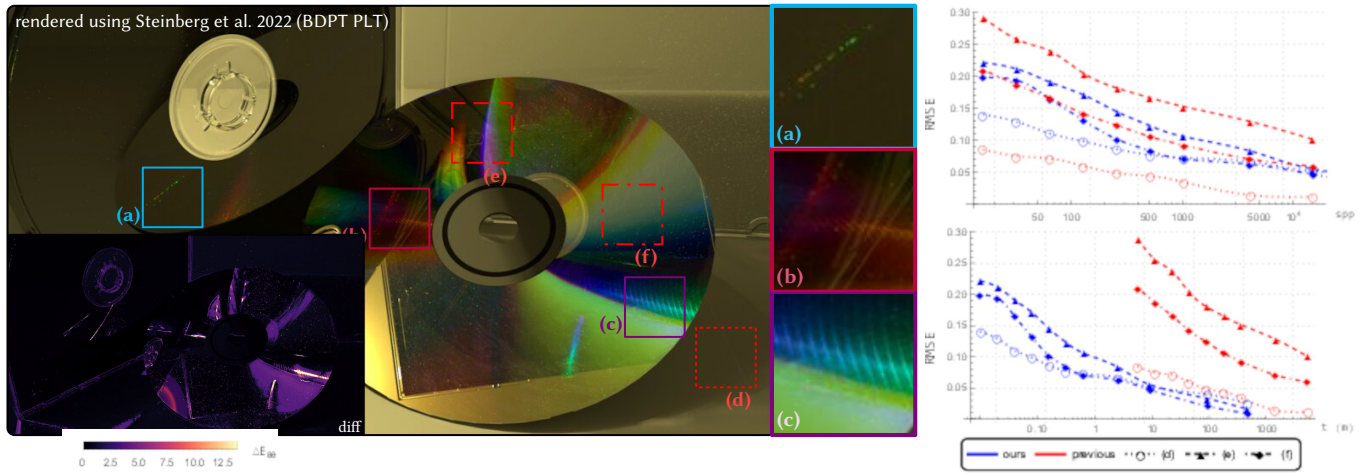

%% file: fig_snake.tex
\begin{figure*}[t]%
    \centering
    \tikzset{external/optimize=true}%
    \tikzsetnextfilename{snake}%
    \resizebox{1\linewidth}{!}{\begin{tikzpicture}[
        ]
    \end{tikzpicture}}
    \Description{}%
    \caption{
        \textbf{Diffractive materials under different illumination conditions.}
        The appearance of diffractive materials depends on light's spectral, polarimetric and coherence properties.
        (a) During daytime, the illumination reaching the snake is dominated by indirect sunlight.
        At some angles, where incident light has a narrower angular spread (and thus is more coherent), the diffractive scales show clear interference patterns.
        (b) At night, direct light from the fluorescent lamps is the primary source of illumination, however that light is too incoherent to produce visible diffraction patterns.
    }%
    \label{fig_snake}
\end{figure*}
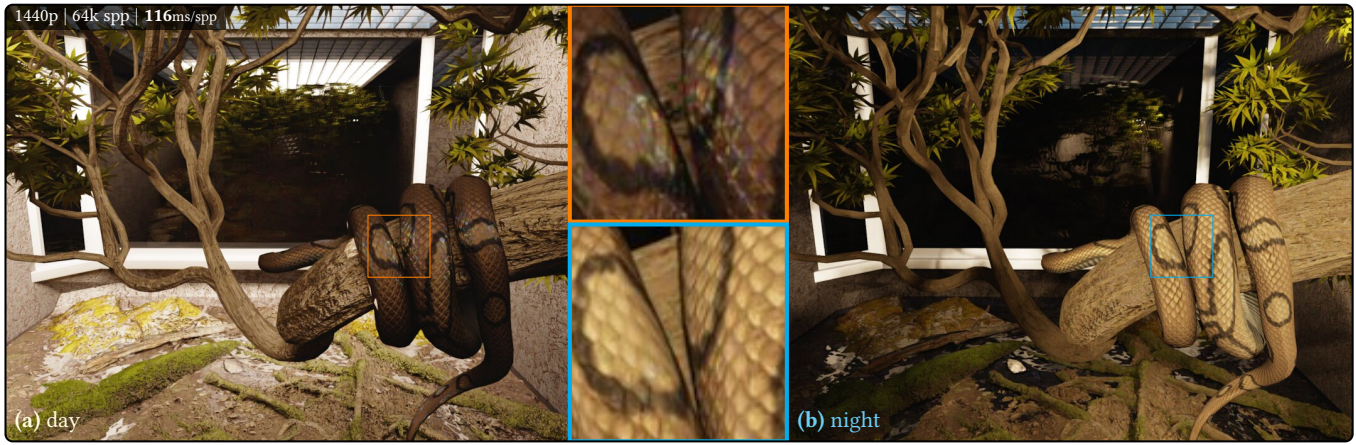